%% file: Ripple_arxiv19.tex
\documentclass[pageno]{jpaper}

\usepackage[normalem]{ulem}

\usepackage{graphicx}
\graphicspath{{/}}
\usepackage{listings}
\usepackage{xcolor}
\definecolor{olive}{rgb}{0.33, 0.42, 0.18}
\usepackage{makecell}
\usepackage{multirow}
\usepackage{array}
\newcolumntype{C}[1]{>{\centering\let\newline\\\arraybackslash\hspace{0pt}}m{#1}}
\newcolumntype{L}[1]{>{\let\newline\\\arraybackslash\hspace{0pt}}m{#1}}
\usepackage[ruled]{algorithm2e}
\usepackage{epsfig}
\usepackage[table]{xcolor}
\usepackage{fixltx2e}
\usepackage{verbatim}
\usepackage[export]{adjustbox}
\usepackage{subfig}
\usepackage{booktabs}
\usepackage{enumitem}
\usepackage{pifont}
\usepackage{balance}
\usepackage{wrapfig}
\usepackage{lipsum}

\newcommand{\skiourakisize}{\fontsize{9pt}{9pt}\selectfont}

\usepackage{filecontents}

\usepackage{cite}

\hypersetup{colorlinks,linkcolor=blue,citecolor=blue,urlcolor=blue,filecolor=blue}

\newcommand{\smallcapital}{\fontsize{9pt}{10pt}\selectfont}

\begin{document}

\title{Ripple: A Practical Declarative Programming Framework for Serverless Compute}

\author{Shannon Joyner$^1$, Michael MacCoss$^2$, Christina Delimitrou$^1$, and Hakim Weatherspoon$^1$\\$^1$Cornell University, $^2$University of Washington\\
$^1$\{sj677,delimitrou,hw228\}@cornell.edu, $^2$maccoss@uw.edu}

\date{}
\maketitle

\thispagestyle{empty}

\input{abstract}

\input{introduction}

\input{background}

\input{design}

\input{implementation}

\input{applications}

\input{methodology}

\input{evaluation}

\input{related_work}

\input{conclusion}
\section*{Acknowledgements}

We sincerely thank Robbert van Renesse, Zhiming Shen, Shuang Chen, Yanqi Zhang, Yu Gan, and Neeraj Kulkarni for 
their feedback on earlier versions of this manuscript. This work was partially supported by NSF grant CNS-1704742, 
NSF grant CNS-1422088, a Facebook Faculty Research Award, a VMWare Research Award, a John and Norma Balen Sesquicentennial Faculty Fellowship, 
and generous donations from GCE, Azure, and AWS.  
\balance
\bibliographystyle{plain}
\bibliography{bibliography}

\end{document}

%% file: abstract.tex
\begin{abstract}
Serverless computing has emerged as a promising alternative
to infrastructure- (IaaS) and platform-as-a-service (PaaS)
cloud platforms for applications with ample parallelism
and intermittent activity. Serverless promises greater
resource elasticity, significant cost savings,
and simplified application deployment. All major cloud providers,
including Amazon, Google, and Microsoft, have introduced serverless to their public cloud offerings.
For serverless to reach its potential, there is a pressing need
for programming frameworks that abstract the deployment complexity away from the user.
This includes simplifying the process of writing
applications for serverless environments, automating task and data
partitioning, and handling scheduling and fault tolerance.

We present Ripple, a programming framework designed to specifically
take applications written for single-machine execution and allow them
to take advantage of the task parallelism of serverless.
Ripple exposes a simple interface that users can leverage
to express the high-level dataflow of a wide spectrum of applications,
including machine learning (ML) analytics, genomics, and proteomics. 
Ripple also automates resource provisioning, meeting user-defined QoS targets, 
and handles fault tolerance by eagerly detecting straggler tasks. 
We port Ripple over AWS Lambda and show that, across a set of diverse applications, 
it provides an expressive and generalizable programming
framework that simplifies running data-parallel applications on serverless, 
and can improve performance by up to 80x compared to IaaS/PaaS clouds for similar costs. 
\end{abstract}

%% file: introduction.tex
\section{Introduction}
\label{sec:introduction}

An increasing number of popular applications are hosted on public clouds~\cite{BarrosoBook,Gan19,Delimitrou14,Barroso11,kozyrakismicro}. 
These include many interactive, latency-critical services with intermittent 
activity, for which the current Infrastructure-as-a-Service (IaaS) or 
Platform-as-a-Service (PaaS) resource models result in cost inefficiencies, 
due to idle resources. Serverless computing has emerged over the past few years 
as a cost-efficient alternative for such applications, with the added benefit 
that the cloud provider handles all data management, simplifying deployment 
and maintenance for the end user. Under a serverless framework, the cloud provider 
offers fine-grained resource allocations to users who only pay a small amount 
for them when a task is executing, resulting in much lower hosting costs.

Several cloud providers have introduced serverless offerings to their public cloud models, 
including AWS Lambda~\cite{lambda}, Google Functions~\cite{googlefunctions}, and
Azure Functions~\cite{azurefunctions}.
In all cases, with small variations, a user launches one or more ``{\it functions}'', 
which the provider maps to one or more machines. Functions take less than a second 
to spawn, and most providers allow the user to spawn hundreds of functions in parallel.
This makes serverless a good option for extremely parallelizable tasks, or for services 
with intermittent activity and long idle periods.

Big data, scientific, and certain classes of machine learning (ML) analytics are good 
candidates for serverless, as they tend to have ample task parallelism, and 
their storage and compute requirements continue to increase exponentially~\cite{bigdata, datagrowth, hogwild, gibbssampling}.
Hosting such services on traditional IaaS and PaaS clouds incurs significantly 
higher costs compared to serverless, since instances need to be maintained for long periods of 
time to ensure low start-up latencies. Alternatively, if optimizing for cost, tasks from large jobs
are subject to long queueing times until resources become available, despite tasks being ready to execute.
When the input load is bursty the reverse is also true, 
with provisioned resources remaining idle during periods of low load.
While cloud providers have auto-scaling systems in place to provide some elasticity
in the number of allocated instances as load fluctuates~\cite{ec2_autoscale},
elasticity is offered at instance granularity. Additionally, scaling out is 
not instantaneous, resulting in unpredictable or degraded performance, which is especially 
harmful for short-running tasks. 

The premise of serverless is improving \textit{resource elasticity}, and 
\textit{cost efficiency}. To realize this we need programming frameworks
that can (i) extract parallelism from existing cloud applications, (ii) 
simplify writing new applications for serverless compute, (iii) manage 
data dependencies between functions transparently to the user, and (iv) automatically 
handle resource elasticity, data partitioning, fault tolerance, and task scheduling to mask the 
performance unpredictability public clouds are subject to~\cite{Delimitrou16,Delimitrou17,GoogleTrace,Ristenpart09,Bakshi10,Zhang12,Zhang14,Zhang11,Delimitrou19}. 

Serverless frameworks additionally need to be programmable enough 
for users to express generic applications at a high-level, and
flexible enough to circumvent the limitations of current serverless 
offerings, such as AWS Lambda, which limits each function to 512MB of disk space and a 15 minute runtime.
Existing analytics and ML frameworks, like Hadoop~\cite{hadoop}, Spark~\cite{spark}, and TensorFlow~\cite{tensorflow}, 
can handle some of these requirements, but need to run over an active cluster by default.
This is problematic in the case of serverless frameworks where resources are frequently allocated 
and reclaimed, requiring the framework to be re-deployed across runs.
Additionally, such general-purpose frameworks are storage-demanding, in terms of 
both memory and disk, consuming a large fraction of the limited system resources allocated 
per serverless function.

\begin{wrapfigure}[13]{r}[\dimexpr\columnwidth+\columnsep\relax]{12cm}
	\vspace{-0.1in}
	\includegraphics[scale=0.364,trim=0.6cm 0 -0.8cm 6cm,clip=true]{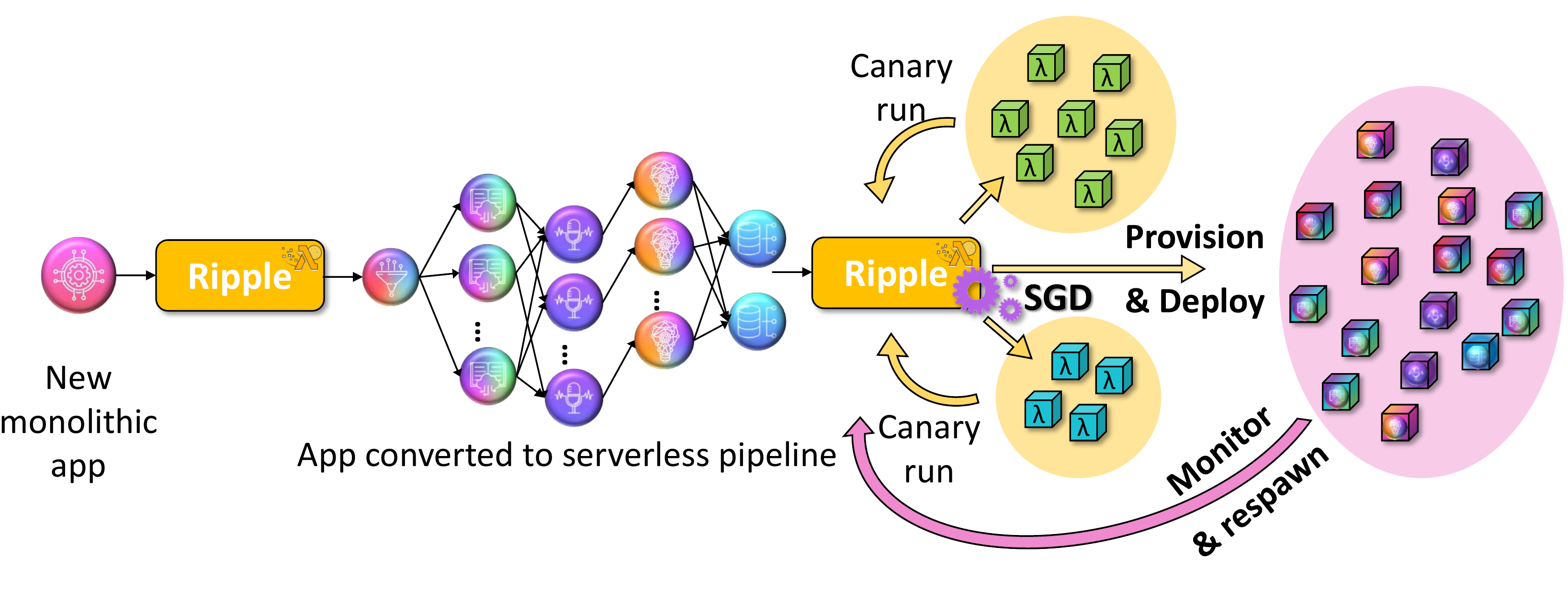}\hspace*{1.4cm}
	\vspace{-0.24in}
	\caption{\label{fig:ripple_overview} Overview of Ripple's operation, from converting a monolithic application to serverless, to profiling and deploying the service to AWS Lambda. }
\end{wrapfigure}
We present \textit{Ripple}, 
a programming framework
designed specifically for serverless computing platforms that run 
highly data-parallel applications. 
Ripple exposes a high-level declarative programming interface 
that enables programmers to express the inherent parallelism in their computation, 
as well as phases that must be executed serially. The framework is general enough to 
capture execution patterns of a wide range of analytics and scientific applications, 
including ML analytics and bioinformatics. 
Ripple uses serverless for all computation phases, 
removing the need for maintaining long-term instances. It also
handles work partitioning automatically, assigning input data chunks 
to serverless functions, and synchronizing them when required. Ripple also allows 
users to express priorities and deadlines for their jobs, implementing a variety 
of scheduling policies, including round-robin, FIFO, and deadline-based scheduling. 
Furthermore, Ripple leverages information about previously-seen jobs to automatically 
infer the degree of concurrency (number of Lambdas per phase) that will allow a new job 
to meet its QoS requirements. Finally, it handles fault tolerance transparently 
to the user, respawning failing or under-performing tasks.


We evaluate Ripple using AWS Lambda. The design of the framework is portable across serverless 
frameworks, including Google Functions and Azure Functions, with minimal modifications.
We focus on three large-scale cloud frameworks, a proteomics framework, DNA compression,
 and a kNN-based classification service used to identify buildings in satellite imaging data~\cite{tide, methcomp, spacenet}.
We show that Ripple offers a simple way for users to express the pipelines of complex applications in a few lines of code,
and that serverless computing offers dramatic performance improvements for data-parallel applications, especially as the size of their datasets increases.
Specifically, we show that Ripple can be up to $80\times$ faster than an EC2 cluster, 
and 30\% cheaper than PyWren~\cite{pywren} for a similar runtime.
We also demonstrate that Ripple automatically handles the resource provisioning of a given job meeting its QoS target, 
and can support different scheduling and fault tolerance mechanisms that further improve performance predictability. 
Ripple is open-source software under a GPL license.

\section{Background}
\label{sec:background}

\subsection{Advantages of Serverless}

Serverless allocates fine-grained resources for a limited time, which reduces the overprovisioning problem current datacenters
experience~\cite{GoogleTrace, Lo15, Delimitrou13, Delimitrou13d, Delimitrou14b, Delimitrou15, Delimitrou14, Delimitrou16}. It also simplifies cloud management; users
simply upload the code they wish to execute to the cloud, and cloud providers handle the state management and resource provisioning.
If an application has ample parallelism, 
\vspace{0.14in}

%% file: background.tex
\vspace*{12\baselineskip}
\noindent serverless achieves much better performance for comparable cost to traditional PaaS or IaaS platforms~\cite{xcamera,gg,salsify}.


Users only pay when a serverless function is running, making the model well-suited for 
applications with intermittent activity. 
Traditional cloud instances can be rented and freed up on-demand, but initializing 
a new instance takes at least 30s. In comparison, spawning and destroying a serverless 
function takes a few milliseconds. For short-running tasks that complete in a few seconds, 
the overhead of instantiating and releasing cloud instances can have a dramatic impact on execution latency.
Some of the most popular serverless computing offerings at the moment include AWS Lambda,
Google Functions, Azure Functions, OpenLambda, and OpenWhisk~\cite{lambda, azurefunctions, googlefunctions, openlambda, openwhisk}. 

\begin{table*}
\begin{tabular}{ p{1.7cm}|p{5.2cm}|p{9.4cm}  }
 \hline
 \multicolumn{3}{c}{\bf{Ripple Programming Interface}} \\
 \hline
 \hline
 &\bf{Description}&\bf{Arguments}\\
 \hline
 {\bf{\fontseries{b}\small\ttfamily{split}}}       &Split a file into small data chunks
&\parbox{9.4cm}{
\vspace*{3px}
\multirow{2}{*}{{{\color{olive}\skiourakisize\fontseries{sb}\small\ttfamily{split\_size}}}: \textit{(optional)} Number of lines per chunk; default 1MB.}\\
}\\
\hline
{\bf\fontseries{b}\small\ttfamily{combine}}     &Combine multiple files
&\parbox{9.4cm}{
\vspace*{3px}
{\color{olive}\fontseries{sb}\small\ttfamily{identifier}}: \textit{(optional)} Field to sort resulting items by.
\vspace*{3px}
}\\
\hline
{\bf{\fontseries{b}\small\ttfamily{top}}}         &Return the top items in a file
&\parbox{9.4cm}{
\vspace*{3px}
{\color{olive}\fontseries{sb}\small\ttfamily{identifier}}: Field to sort items by.\\
{\color{olive}\fontseries{sb}\small\ttfamily{number}}: Number of items to return.
\vspace*{5px}
}\\
\hline
{\bf\fontseries{b}\small\ttfamily{match}}       &Match files to {\fontseries{b}\small\ttfamily i} keyword
&\parbox{9.4cm}{
\vspace*{3px}
{\color{olive}\fontseries{sb}\small\ttfamily{find}}: Property to look for; e.g., highest score sum.\\
{\color{olive}\fontseries{sb}\small\ttfamily{identifier}}: Field to sort items by.
\vspace*{3px}
}\\
\hline
{\bf\fontseries{b}\small\ttfamily{map}}       &Map each item to an input
&\parbox{9.4cm}{
\vspace*{3px}
{\color{olive}\fontseries{sb}\small\ttfamily{input\_key}}: Key parameter to set input name to for next phase.\\
{\color{olive}\fontseries{sb}\small\ttfamily{map\_table}}: Table containing list of files to map.\\
{\color{olive}\fontseries{sb}\small\ttfamily{table\_key}}: Key parameter to set table name to for next phase.\\
{\texttt{\color{olive}\fontseries{sb}\small\ttfamily{directories}}}: \textit{(optional)} Use table directories to map.
\vspace*{3px}
}\\
\hline
{\bf\fontseries{b}\small\ttfamily{sort}}        &Sort file using Radix sort
&\parbox{9.4cm}{
\vspace*{3px}
{\color{olive}\fontseries{sb}\small\ttfamily{identifier}}: Field to sort items by.
\vspace*{3px}
}\\
\hline
\multirow{2}{*}{\bf\fontseries{b}\small\ttfamily{partition}}      &Return {\fontseries{b}\small\ttfamily n} equally spaced ranges (used for Radix sort)
&\parbox{9.4cm}{
\vspace*{3px}
\multirow{2}{*}{{\color{olive}\fontseries{sb}\small\ttfamily{identifier}}: Field used to partition the items into ranges.} 
\vspace*{3px}
}\\
\hline
{\bf\fontseries{b}\small\ttfamily{run}} &Invoke Lambda function
&\parbox{9.4cm}{
\vspace*{3px}
{\color{olive}\fontseries{sb}\small\ttfamily{application}}: The name of the application file to execute.\\
{\texttt{\color{olive}\fontseries{sb}\small\ttfamily{output\_format}}}: \textit{(optional)} Format of the output file.
\vspace*{3px}
}\\
\hline
\end{tabular}
\caption{The programming interface of Ripple. }
\label{table:functions}
\end{table*}

\subsection{Limitations of Serverless}

Despite its increasing popularity, serverless still needs to overcome several system challenges. 
First, while cloud providers make triggering serverless functions easy, they do not currently offer 
an easy way to share state between functions. If a user wants to combine results
from multiple functions, or save state to use later in the pipeline, this state needs to be saved 
in remote storage -- S3 for AWS Lambda -- potentially degrading application performance~\cite{lambda,step}. 
Second, current serverless functions are resource-constrained, typically being limited to a single CPU, a couple 
GB of memory, and a few hundred MB of disk space. This limits the type of computation that can be hosted 
on serverless, especially given the fact that current platforms
do not offer a way for users to partition their work to functions in a programmable and systematic way.
This is especially challenging for non-expert cloud users running applications with complex workflows,
where the degree of available parallelism varies across execution phases.
Third, despite the increased visibility cloud providers have into serverless applications as opposed to 
traditional PaaS- or IaaS-hosted applications, current platforms still do not have a way for 
users to express a quality-of-service (QoS) target for their applications that the cloud provider must meet. 
This is especially detrimental in serverless frameworks where resource sharing, even of a single CPU, is 
much more prevalent than in traditional cloud systems, and hence applications are more prone to unpredictable 
performance due to interference. 

The challenges and opportunities of serverless put increased pressure
on programming frameworks that simplify porting traditional cloud
applications on to serverless platforms in a way that masks their current limitations.

%% file: design.tex
\section{Ripple Design}
\label{sec:design}

We propose Ripple, a general-purpose programming framework for serverless that
targets applications with ample parallelism. Ripple follows the model of a high-level 
declarative language; a user expresses a job's dataflow at a high level, and Ripple abstracts away 
the complexity of data partitioning (Sec.~\ref{sec:functions}), 
task scheduling (Sec.~\ref{sec:scheduling}), resource provisioning (Sec.~\ref{sec:provisioning}), 
and fault tolerance (Sec.~\ref{sec:fault_tolerance}). 
Fig.~\ref{fig:ripple_overview} shows the high-level overview of the framework. 

\subsection{Ripple Programming Framework}
\label{sec:functions}

Ripple exposes a concise programming interface of eight principal functions 
that users can leverage to express the most common parts of their services' dataflow,
including parallel and serial phases, and dependencies between dataflow stages. 
Each function can be used in one or more stages of a job. The programming 
interface is shown in Table~\ref{table:functions}. Apart from the eight functions 
in the table, users can also upload arbitrary operations, 
and invoke them through the {\fontseries{sb}\small\ttfamily run} function. 

\lstdefinestyle{customcpp}{
aboveskip=0in,
belowskip=0in,
abovecaptionskip=0.08in,
belowcaptionskip=0in,
captionpos=b,
xleftmargin=\parindent,
language=C++,
morekeywords={import,config,pipeline,step,in,sort,run,compile},
showstringspaces=false,
basicstyle={\linespread{0.6}\fontseries{sb}\small\ttfamily},
keywordstyle=\bfseries,
commentstyle=\itshape\color{green!40!black},
}

\begin{figure}[h]
\vspace{-0.03in}
\begin{lstlisting}[style=customcpp, caption=Ripple Configuration for DNA Compression., label=lst:dna] PrioQueue<Time, GateInput> eventQueue;
import ripple
// Configure region and Lambda resources
config = {
  "region": "use-west-2", 
  "role": "aws-role", 
  "memory_size": 2240, 
}
// Express computation phases
pipeline = ripple.Pipeline(
  name="compression", 
  table="s3://my-bucket", 
  log="s3://my-log", 
  timeout=600, 
  config=config
)
// Define input data
in = pipeline.input(format="new_line")
// Declare how to sort input data
step = input.sort(
  identifier="start_position",
  params={"split_size": 500*1000*1000}, 
  config={"memory_size": 3008}
)
// Declare compression method 
step = step.run(
  "compress_methyl",
  params={"pbucket": "s3://my-program"}
)
// Create and upload functions
pipeline.compile("json/compile.json")

\end{lstlisting}
\vspace{0.05in}
\end{figure}

Algorithm~\ref{lst:dna} shows an example Ripple application for porting a simple DNA compression algorithm to AWS Lambda.
The \texttt{config} parameter contains the basic setup parameters, 
such as the default amount of memory to use for functions.
The {\tt pipeline} is initialized by specifying the job configuration 
and the result and logs locations. Next, an input variable is declared.
The input variable declares the data format of the input dataset, 
which helps Ripple determine how to split and manipulate the data.
The input DNA files can be gigabytes in size, significantly more than 
the 512MB of disk space or 3GB of memory space offered by Lambda.
The compression algorithm relies on finding sequences with similar prefixes.
Therefore, Ripple calls \texttt{sort} on the input file, which sorts the file into 500MB chunks.
The user can also provide an application-specific hint on the split size a function should handle.
In this example, \texttt{sort} needs more than 2240MB, so the memory allocation is 3008MB instead.
The next step compresses the input shards, invoked via Ripple's \texttt{run} function. 
The application function returns the paths of the files that Ripple should pass to the next pipeline step.

Once the user specifies all execution steps, they can compile the pipeline, 
which generates the {\smallcapital JSON} file Ripple uses
to set up the serverless functions. Ripple then uploads the code 
and dependencies to AWS, and schedules the generated tasks.

\subsection{Automating Resource Provisioning}
\label{sec:provisioning}

One of the main premises of serverless is simplifying resource provisioning, by letting the cloud provider handle resource allocation and elastic scaling. 
Serverless frameworks today still do not achieve that premise, by requiring the user to specify the number of serverless functions a job is partitioned across. 
In addition to allowing users to specify priorities among their jobs (Sec.~\ref{sec:scheduling}), 
Ripple also automates the resource provisioning and scaling process for submitted applications. 
Specifically, when a new application is submitted to AWS Lambda via Ripple, the user also expresses a deadline that the job needs to complete before~\cite{Zaharia10,Srinivasan02}. 
Ripple uses that deadline to determine the right resources for that job, depending on whether the job consists of one or multiple phases. 

\noindent{\bf{Single-phase jobs: }}This is the simplest type of serverless jobs, where the input data is partitioned, processed, and the result is combined, 
and either stored on S3 or sent directly to the user. In this case, Ripple just needs to determine the number of Lambdas needed for processing. 
Ripple uses a two-step process to determine the appropriate degree of concurrency. First it selects a small partition of the original input dataset, 
{\fontseries{sb}\small\ttfamily min(20MB,full\_input)}, called \textit{canary input}.~\footnote{The 20MB lower limit
is determined empirically to be sufficient for our applications, and can be tuned for different services. } By default Ripple selects 
the canary input starting from the beginning of the dataset, unless otherwise specified by the user. It then uses the canary input 
as the input dataset for two small jobs, each with different data split sizes per Lambda. 
One job runs with the default split size for AWS Lambda, 1MB, and the other with the split size Lambdas 
would have if the job used its maximum allowed concurrency limit (1,000 on AWS by default). 
To track task progress when enforcing different resource allocations, Ripple implements a simple tracing system detailed in Sec.~\ref{sec:implementation}. 
Ripple's tracing system collects the per-Lambda and per-job execution time (including the overheads for work partitioning and result combining), 
and uses this information to infer the performance of the job with any split size. 
Inference happens via Stochastic Gradient Descent (SGD), where SGD's input is a table with jobs as rows and split sizes as columns. SGD has been shown to 
accurately infer the performance of applications on non-profiled configurations, 
using only a small amount of profiling information~\cite{Klimovic18a,Klimovic18,Delimitrou13,Delimitrou14}, 
in the context of cluster scheduling, heterogeneous multicore configuration, and storage system configuration. 
As new jobs are scheduled by Ripple, the number of rows in the table increases. 
Given that input dataset sizes varies across jobs, the table includes 
profiling information on a diverse set of split sizes, including one split size 
(default=1MB) that is consistent across all jobs. 

Once Ripple determines the level of concurrency that will achieve a job's specified deadline, 
it launches the full application and monitors its performance. If measured performance deviates from the estimated 
performance with that degree of concurrency, Ripple updates the table with 
the measured information to improve prediction accuracy over time. 

Apart from specifying per-job deadlines, users can also request that Ripple tries to achieve 
the best possible performance for a job, 
given AWS's resource limits, 
or alternatively, they can specify an upper limit for the cost they are willing to spend for a job, 
in which case Ripple maximizes performance under that cost constraint. 


\noindent{\bf{Multi-phase jobs: }}Typical serverless applications involve multiple phases 
of high concurrency, which may be interleaved with sequential phases that combine 
intermediate results. The process Ripple follows to provision such jobs is similar to 
the one described above, with the difference that columns in the SGD table are now ratios 
of concurrency degrees across phases. Ripple also needs to increase the number of canary runs 
to ensure high prediction accuracy from two for single-phase jobs to four for multi-phase jobs. 
Split sizes per phase are selected in [1MB,fullDataset/maxLambdas] as before. If measured 
performance deviates from estimated, Ripple again updates the corresponding column in the table 
to improve accuracy the next time a similar job is scheduled. 

Note that higher Lambda concurrency does not necessarily translate to better performance for three reasons. 
First, getting a speedup as Lambdas increase is contingent on the available parallelism in a job's computation. 
Applications with a low inherent parallelism will not benefit from a higher degree of concurrency, 
while also incurring higher costs for the increased number of serverless functions. 
Second, parallelism does not come for free, as work and data needs to be partitioned 
and distributed across tasks, and results need to be reassembled and combined before 
computation can proceed. This is especially the case for multi-stage job pipelines, 
where parallel phases are interleaved with phases that combine intermediate results. 
Third, a larger number of concurrent serverless functions also increases the probability 
that some of these tasks will become stragglers and degrade the entire job's execution time. 

Obtaining a small amount of profiling information before provisioning a new job allows 
Ripple to identify if either of the first two reasons that would prevent a job from benefiting from higher 
concurrency are present. The third reason is impacted both by job characteristics, 
e.g., faulty data chunks can cause some Lambdas to become stragglers, and 
by the state of the serverless cluster, e.g., some Lambdas can underperform due 
to interference between tasks sharing system resources. To address underperforming 
tasks, we also implement a fault tolerance mechanism in Ripple, detailed below. 

\subsection{Fault Tolerance}
\label{sec:fault_tolerance}

Task failures are common in data-parallel analytics jobs. 
Furthermore, straggler tasks are a well-documented occurrence, 
where a small number of under-performing tasks delay the completion 
of the entire application~\cite{Zaharia08,Dolly,tarazu,Mantri10,Lin09,Gandhi11}.
AWS Lambda does not provide handlers to running functions, 
therefore a function's progress cannot be directly monitored.
Instead, Ripple collects execution logs for each Lambda request based on when they write to S3. 
These logs do not only prevent duplicate requests,
but contain payload information to re-execute failed Lambda processes.
If Ripple detects that a function has not been logged within the 
user-specified timeout period from its instantiation, it will re-execute 
that function. This also helps minimize the impact of stragglers, 
as Ripple can eagerly respawn tasks that are likely to under-perform. 

\subsection{Scheduling Policies}
\label{sec:scheduling}

Cloud users typically submit more than one job to a public cloud provider. 
{\textsc AWS} Lambda at the moment implements a simple 
{\textsc FIFO}\footnote{Launching Lambdas happens in a ``mostly'' FIFO order, however the exact ordering is not strictly enforced. } 
scheduling policy, which means that Lambdas submitted first will also start 
running first. This does not allow a lot of flexibility, especially when 
different jobs have different priorities or must meet execution time deadlines.

Ripple implements several scheduling policies users can select when they submit their applications. 
To prevent conflicts between scheduling policies specified in different jobs, a scheduling policy applies
to all active jobs managed by Ripple. 
Ripple precomputes the best task schedule based on the specified scheduling policies, 
and enforces scheduling policies by reordering tasks client-side, 
and deploying them in that order on AWS Lambda. 
Below we summarize each scheduling policy supported in Ripple. 

\noindent\textbf{FIFO: } This is the default scheduling policy implemented by AWS Lambda.
Functions are executed in the order of submission. Under scenarios of high load, FIFO can lead to long queueing latencies.

\noindent\textbf{Round-Robin: } Ripple interleaves the phases of each job to allocate approximately
equivalent time intervals to each application. Round-Robin penalizes the
execution time of the first few jobs, but reduces queueing delays, and improves fairness compared to FIFO.

\noindent\textbf{Priorities: } Priorities in Ripple are defined at job granularity, 
with high-priority jobs superseding concurrently-submitted low-priority jobs. 
If a high priority job arrives when the user has reached their resource quota on 
Lambda (1,000 active Lambdas by default), Ripple will pause low priority jobs, 
schedule the high priority application, and resume the former when the high 
priority job has completed. Applications of equal priorities revert to Round-Robin.
Unless otherwise specified, we use the {\textsc FIFO} scheduler by default. 

%% file: implementation.tex
\section{Implementation}
\label{sec:implementation}

Ripple is written in approximately 5,000 lines of code in Python,
including application specific code and configurations. 
The eight functions shown in Table~\ref{table:functions}
are also written in Python. A user can create arbitrary functions, 
as long as the functions adhere to Ripple's \texttt{run} template.
Currently, Ripple supports interactions with AWS Lambda and S3.
However, Ripple's API has been designed to be abstractable to other 
providers, including Windows Azure and Google. 

\noindent\textbf{Job configuration: }A user ports a new application 
by using the Ripple API to create the application pipeline. This pipeline 
is compiled to a {\textsc JSON} configuration file. The configuration file 
indicates which Lambda functions to use and the order in which to execute the functions.
If the application requires manipulation of file formats that the framework 
does not support by default\textemdash such as newline separated, TSV, FASTA, 
mzML)\textemdash, users simply need to implement a format file for the framework 
to know how to parse the input files. Users also need to create two tables: one 
to store the input, intermediate, and output files, and another for the log files 
Ripple maintains to manage the execution of functions.

To deploy an application, the user follows an automated setup process in Ripple, 
which takes their configuration file and uploads all necessary code to AWS.
The maximum deployment package size AWS currently allows is 50MB. However, AWS 
provides layers for users to import extra dependencies. Ripple checks the user's 
program for a list of common dependencies, such as {\tt numpy} or {\tt scikit-learn},
and links the functions to their corresponding dependencies~\cite{numpy, sklearn}.
If the layer does not exist, the user can create and add the new layer to Ripple.
Ripple does not use any additional containerization or virtualization processes 
beyond what AWS Lambda implements by default. 

\vspace{0.05in}
\noindent\textbf{Tracing and monitoring: }Ripple needs to monitor the execution of 
active functions to adhere to the specified scheduling policies, and ensure performant 
and fault tolerant operation. Since AWS Lambda does not provide handlers to active 
functions, Ripple periodically polls S3 to see if a function has finished, and records 
the function instantiation and completion events in its execution log. The asynchronous 
nature of Lambda makes identifying failed functions challenging. For every job, Ripple 
spawns a thread to monitor the job's progress based on its logs. Ripple waits until the 
timeout period specified by the user for a given function's log. If the log does not 
appear, Ripple re-invokes the function. For scheduling algorithms, such as deadline-based 
scheduling, Ripple needs the ability to ``pause'' a job. To enable this, we specify a 
pause parameter for each job that indicates when functions for that job should stop executing.
To re-execute, Ripple re-triggers the pipeline, and checks if the task has already executed.
If so, the function simply re-triggers its child processes.
We have verified that the tracing system has no noticeable impact on either Lambda throughput or latency. 

\vspace{0.05in}
\noindent\textbf{Testing library: } Due to the asynchronous nature of serverless, 
testing can be difficult as logs are spread over multiple files.
Ripple contains logic to make testing code easier. A user can specify files to use 
as source data and Ripple will locally and serially go through each stage of the pipeline.

\vspace{0.05in}
\noindent\textbf{Fault tolerance: } Finally, Ripple itself may fail for a number of reasons,
including network outage, server failure, or internal error. To ensure that such a failure does
not leave active functions unmanaged, we maintain Ripple's execution log in persistent storage,
such that, if a failure occurs, a hot stand-by master can take over management of the service. 

%% file: applications.tex
\section{Applications}
\label{sec:applications}

\begin{table}
\vspace*{3px}
\fontfamily{cmr}\selectfont
\begin{tabular}{lcc}
\hline
{\bf Application}    & {\bf JSON file} & {\bf \texttt{run} functions} \\ 
\hline
\hline
SpaceNet        & 16 & 250 \\
Proteomics            & 25 & 86 \\
DNA Compression & 13 & 36 \\
\hline
\end{tabular}
\caption{Lines of Ripple code for each application. ``JSON file'' shows the LoC for the configuration file, 
and ``\texttt{run} functions'' shows the LoC for the specific logic of the application.}
\label{loc}
\end{table}

We use three applications to demonstrate the practicality and generality of Ripple.
In all cases, the modifications needed to port the original applications to Ripple are minimal.
Ripple exposes a high-level declarative interface, which enables users to take code designed 
to run on a single-machine and port it to a serverless platform. 
Below we describe each of the application frameworks we port on Ripple.
The logic of the protein analytics and DNA compression applications did not require any changes to be ported to Ripple.
The SpaceNet Building Border Identification application was reimplemented in Python, given AWS's languages limitations. 
Table~\ref{loc} shows the Ripple lines of code needed for the Ripple {\smallcapital JSON} configuration file, 
and for the {\tt run} function that is specific to each application's logic.

\subsection{SpaceNet Building Border Identification}

\begin{figure}
\centering
\includegraphics[scale=0.426,viewport=80 0 500 240]{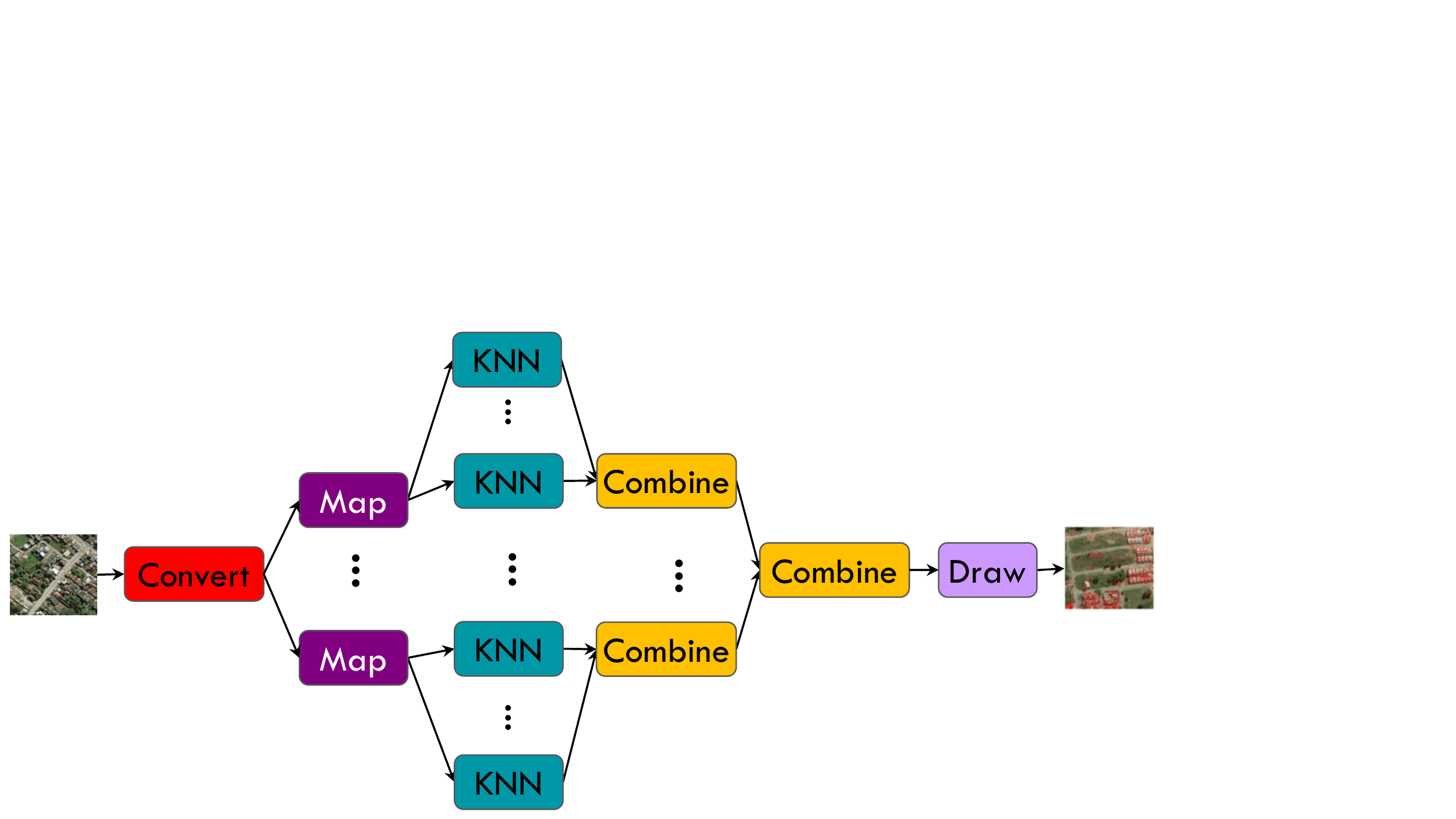}
\caption{Pipeline for SpaceNet. The \texttt{convert} shards the input data. 
\texttt{map} pairs test with training data chunks. The \texttt{KNN} stage performs
a brute force KNN. The first \texttt{combine} stage finds the absolute nearest neighbors
for every pixel, and the second \texttt{combine} stage combines all results into one file.
The last stage colors the identified border pixels.}
\label{fig:spacenet_pipeline}
\end{figure}

SpaceNet hosts a series of challenges, including identifying buildings from satellite 
images~\cite{spacenet}. The developers provide {\textsc TIFF} training and test images 
of different areas, only some of which contain buildings. We implemented our own solution 
for this challenge using a K-Nearest Neighbor (kNN) algorithm~\cite{knn}. For each pixel 
in the training images, we use the SpaceNet solutions to classify the pixel as a border, 
inside or outside the building. In the test image, we identify the most similar pixels 
in the training images, and classify the test pixels accordingly. SpaceNet offers both 
high resolution, 3-band images, and low resolution, 8-band images. For our experiments, 
we used the 3-band training and test images.  
Fig.~\ref{fig:spacenet_pipeline} shows the Ripple pipeline for SpaceNet.

\subsection{Proteomics}
Tide is a protein analytics tool used for analyzing protein sequences~\cite{tide}.
Tide takes experimental data and scores them 
\begin{wrapfigure}[11]{l}{0.274\textwidth}
	\vspace{-0.14in}
  \includegraphics[scale=0.40,trim=0 0 6cm 8.6cm,clip=true]{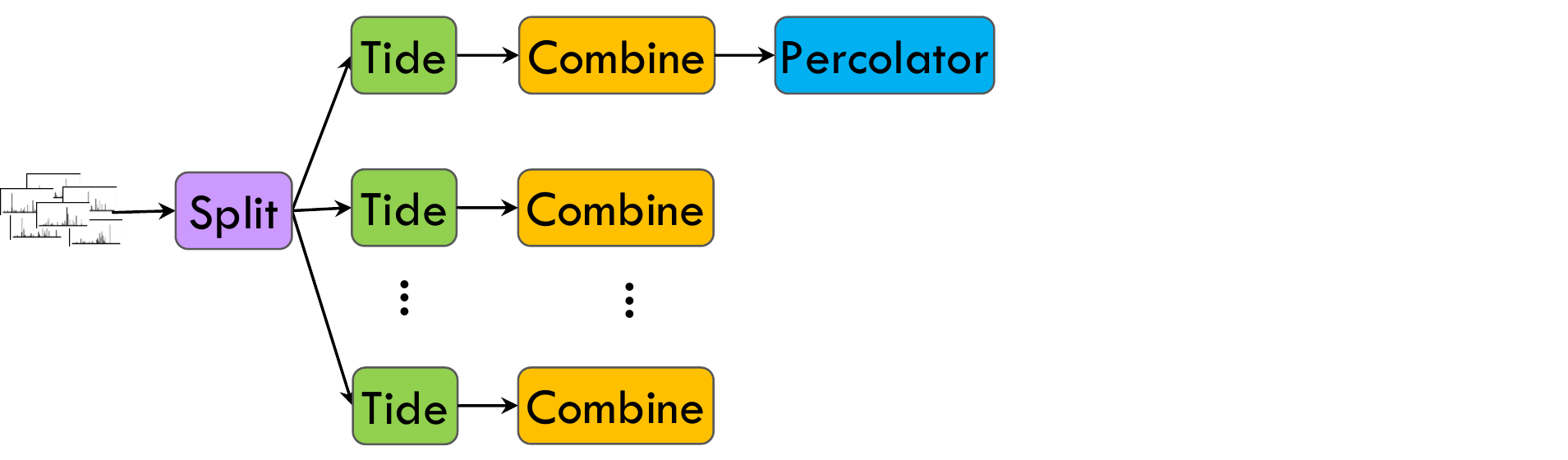}
	\vspace{-0.22in}
  \caption{Proteomics with Tide~\cite{tide}. We first split the input into shards. 
  Next, Tide, determines the protein composition of each chunk. The 
  output is processed by Percolator.}
  \label{fig:tide_lambda_pipeline}
\end{wrapfigure}
against theoretical sequences, 
defined in a file containing protein sequences predicted from the genome sequence~\cite{tide, sequest}.
Experimental sequences are in mzML format, an {\textsc XML}-based format~\cite{mzml, proteowizard}, and theoretical sequences are in {\textsc FASTA} format~\cite{fasta}.
Results from Tide are often given to an ML application called Percolator
to score the confidence of the protein composition identification~\cite{percolator}. 
We evaluate both Tide and Percolator in a single pipeline. Fig.~\ref{fig:tide_lambda_pipeline} 
shows the Ripple pipeline for the proteomics application.

\subsection{DNA Compression}

\begin{figure}
\centering
\vspace{-0.4in}
\includegraphics[scale=0.374,trim=0 0 0.4cm 6.0cm, clip=true]{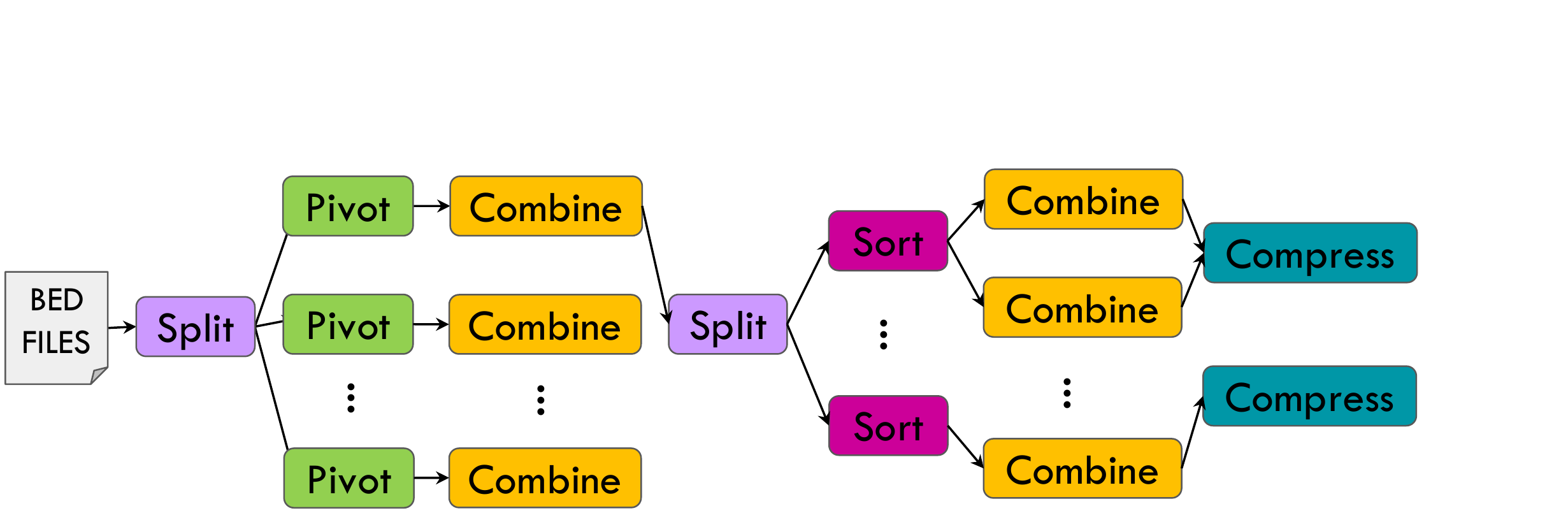}
\vspace{-0.08in}
  \caption{Pipeline for DNA compression. The input BED file is sorted using Radix sort. 
	  This involves finding pivots in chunk segments. Once pivots are found, the file is split, 
  and chunks are compressed in parallel. }
  \label{fig:methyl_lambda_pipeline}
\end{figure}

{\textsc METHCOMP} is an application to compress and decompress {\textsc DNA} methylation data~\cite{methcomp}.
The data from this pipeline are stored in a {\textsc BED} format, which is newline delimited~\cite{bedmethyl}.
This format remains human-readable, but readability comes at the expense of high storage costs.
Nonetheless, {\textsc METHCOMP} is both memory- and disk-intensive 
and the algorithm does not use threads or multiple cores by default.
Figure~\ref{fig:methyl_lambda_pipeline} shows the Ripple pipeline 
for the {\textsc METHCOMP} application. The compression pipeline was modified to sort the input dataset.

%% file: methodology.tex
\section{Methodology}
\label{sec:methodology}

We run each application on AWS Lambda using Ripple.
Unless otherwise specified, we run experiments in regions 
that had a max concurrency of 1,000 functions. We compare 
Ripple to two other setups; AWS EC2 with resource 
autoscaling~\cite{ec2_autoscale}, and PyWren~\cite{pywren}. 

\begin{wrapfigure}[13]{r}{0.30\textwidth}
  \includegraphics[scale=0.352,viewport=20 20 400 300]{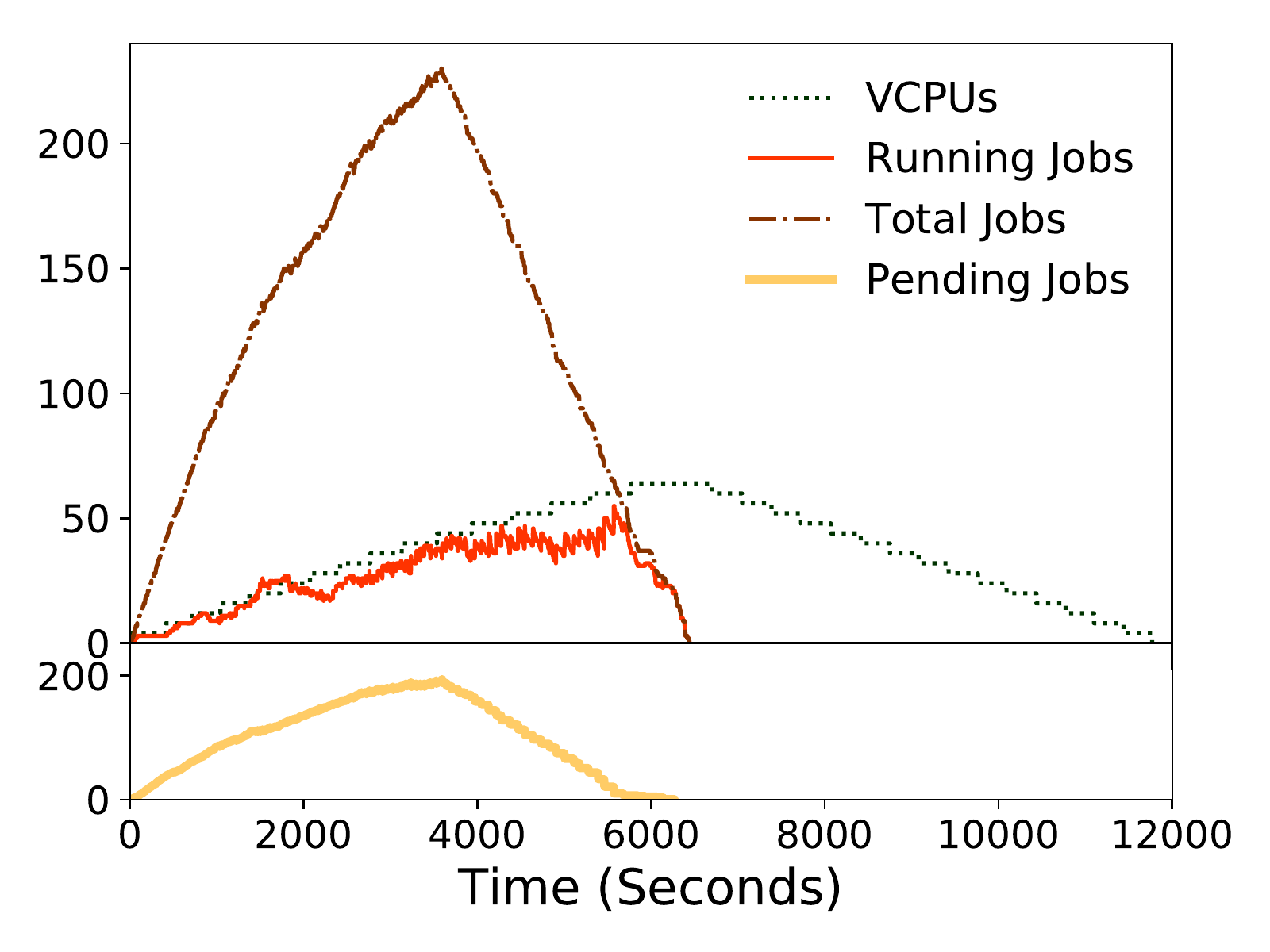}
  \caption{Tide runtime on EC2 for a uniform simulation using the default autoscaling policy (in 5min intervals).}
  \label{fig:tide_uniform_default_comparison}
\end{wrapfigure}

\noindent\textbf{EC2 Autoscaling: }
We use the Amazon Elastic MapReduce (EMR) cluster to compare Ripple to EC2 autoscaling. 
EMR only allows servers to be added if usage for one or more resources (CPU, memory) 
is above a certain threshold for more than five minutes. 
Figure~\ref{fig:tide_uniform_default_comparison} shows how Tide scales 
using EMR's default policy if a job is sent every 10 seconds for an hour. 
A server is added if CPU utilization is above 70\%, 
and removed if CPU utilization is below 30\%. 
The bottom portion of the graph shows the number of pending jobs. 
The default 5 minute scaling policy could not handle rapid load increases, 
taking almost 2 hours to process all requests and an additional 2 hours 
to terminate all servers.

To show that even a more agile version of EC2's autoscaling has suboptimal elasticity compared to Ripple, 
we also implemented a version of EC2 autoscaling that increments and decrements instances at 10s granularity. 
We implemented this policy using the {\tt boto3} API, and ran all jobs inside Docker containers to simplify scaling out~\cite{boto3, docker}.
Since the Docker containers often needed significant disk space, for every EC2 machine, we allocated a 26GB volume of persistent storage. 

\noindent\textbf{PyWren: }PyWren is an open-source programming framework for serverless, and the most closely-related prior work to Ripple~\cite{pywren}. 
	PyWren is designed following the MapReduce style of distributed processing, with mappers running on AWS Lambda 
	as a single map phase, followed by reducers running on traditional EC2 instances. 

\noindent Below we describe each application's setup in Ripple, PyWren, and EC2. 

\subsection{SpaceNet Building Border Identification}

We use 1,000 images to train the border classifier.  
The feature vector for each pixel was the RGB value of the pixel and its 8 surrounding neighbors.
We store these results in an {\textsc S3} bucket.
For each test image, we use the 100 closest neighbors to classify a pixel.
Finally, we color the identified border pixels.
For all implementations, we use the {\tt scikit-learn} library to compute kNN using brute force~\cite{sklearn}.
For SpaceNet, we ran all Lambda experiments in a region where the max concurrency was 5K functions.

\noindent{\bf{Ripple setup: }}
We create a function to convert TIFF testing images to their feature vectors. 
Given the amount of training and testing data, 
we split testing images into subsets of 1,000 pixels, 
and mapped each subset to a subset of training data to run kNN.
This typically resulted in approximately 180 subsets of pixels, 
each paired with around 40 chunks of training data, and a total 
of 7K Lambdas. Results are combined in two phases; the first phase found 
the absolute 100 nearest neighbors for each pixel, 
and the second phase concatenated the partial outputs.
Each pipeline stage is triggered when the output of 
the previous phase is written to S3. 

\noindent{\bf{PyWren setup: }} 
For fairness, we attempted to run all parallel tasks on Lambda; however, 
the first combine step caused out of memory errors on Lambda; 
we subsequently moved the \texttt{combine} tasks to a \mbox{\texttt{r4.16xlarge}} EC2 instance. 

\noindent{\bf{EC2 setup: }} Computing the k nearest neighbors is very 
memory intensive, so we use \texttt{r5a.xlarge} machines.

\subsection{Proteomics}

\noindent{\bf{Ripple setup: }}
Tide frequently consumes more than the max 3GB offered by Lambda.
Therefore, the first step is splitting 
the mzML input file into small chunks. We then run Tide on each file.
Afterwards, we combine the results in a single output file.
For all experiments, we used human FASTA files. 

\noindent{\bf{PyWren setup: }} Tide is both CPU- and memory-intensive, 
therefore we use Lambda for parallel phases, and a \texttt{t2.xlarge} 
EC2 instance for serial computation. 

\noindent{\bf{EC2 setup: }} We again use \texttt{t2.xlarge} instances.

\subsection{DNA Compression}

\noindent{\bf{Ripple setup: }}
We first use Ripple's \texttt{sort} to sort the input
so that similar sequences are near each other, and subsequently compress each data chunk.

\noindent{\bf{PyWren setup: }} We again use Lambda for parallel phases, 
and a \texttt{t2.xlarge} EC2 instance for serial computation. 

\noindent{\bf{EC2 setup: }} As before, we use \texttt{t2.xlarge} instances.

%% file: evaluation.tex
\section{Evaluation}
\label{sec:evaluation}

Below, we answer five questions, pertaining to the performance, cost, fault tolerance, and scalability of Ripple, 
as well as its ability to preserve the QoS requirements of diverse jobs. 

\begin{itemize}
	\item {How able is Ripple's provisioning system to preserve QoS for incoming jobs (Sec.~\ref{sec:evaluation:accuracy})?}
\end{itemize}


\begin{itemize}
	\item {How robust is Ripple across job distributions (Sec.~\ref{sec:evaluation:workloads})?}
\end{itemize}



\begin{itemize}
	\item {How does Ripple compare to PyWren (Sec.~\ref{sec:evaluation:pywren})?}
\end{itemize}



\begin{itemize}
	\item {How does Ripple scale as job concurrency increases (Sec.~\ref{sec:evaluation:concurrency})?}
\end{itemize}


\begin{itemize}
	\item {How does the fault tolerance mechanism in Ripple affect application performance (Sec.~\ref{sec:evaluation:fault_tolerance})?}
\end{itemize}



\begin{figure}
	\centering
	\begin{tabular}{cc}
		\includegraphics[scale=0.20,viewport=210 0 580 430]{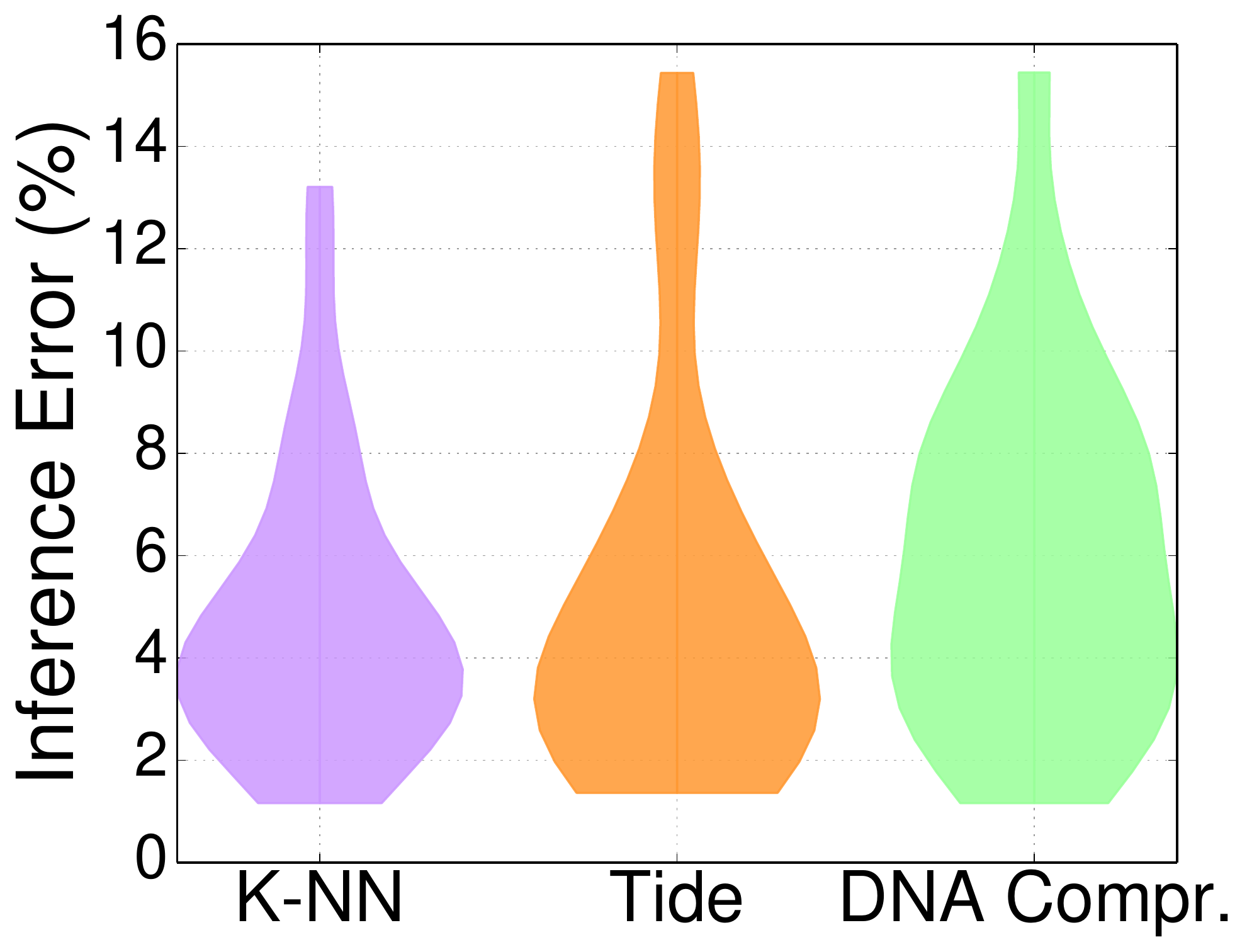} &
		\includegraphics[scale=0.20,viewport=70 0 440 430]{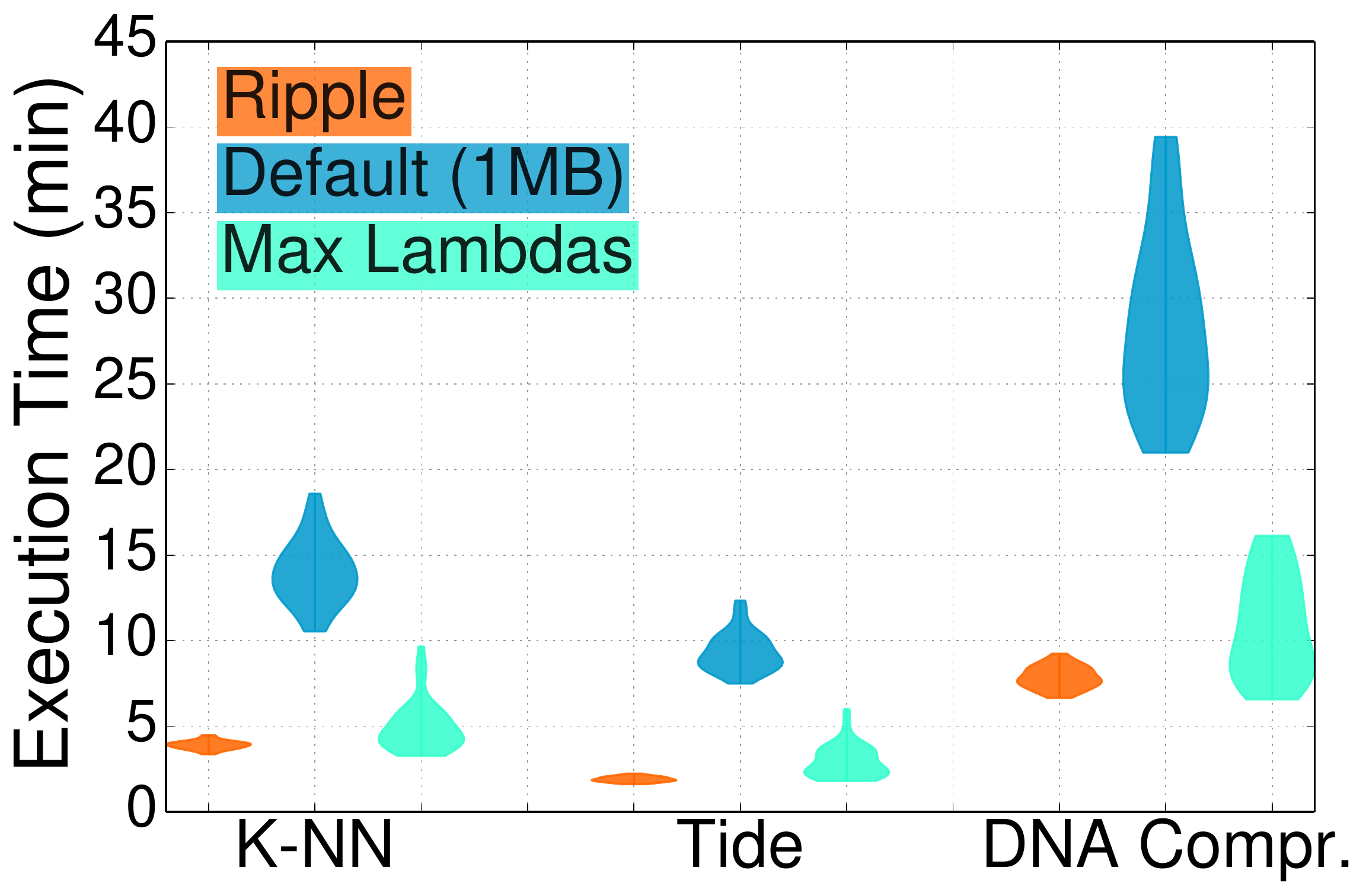} \\
	\end{tabular}
	\caption{(a) Error between execution time estimated by SGD in Ripple across the three applications, and measured execution time with the corresponding 
	resource configuration. (b) Performance with Ripple compared to using the default 1MB data chunks per Lambda, and the maximum number of Lambdas allowed, across 
the three applications. }
	\label{fig:error_perf}
\end{figure}

\begin{table}
	\fontfamily{cmr}\selectfont
	\begin{tabular}{l||ccc}
		\hline
		\multirow{2}{*}{\bf Application}    & \multicolumn{3}{c}{\bf Cost}\\ 
		\cline{2-4}
		& {\bf Ripple} & {\bf 1MB} & {\bf Max Lambdas} \\
		\hline
		\hline
		SpaceNet        & \$2.77 & \$3.81 & \$4.61 \\
		Proteomics      & \$0.42 & \$0.61 & \$0.53 \\
		DNA Compression & \$0.36 & \$0.51 & \$0.42 \\
		\hline
	\end{tabular}
	\caption{Cost across applications for (i) Ripple, (ii) the default split size, and (iii) the max number of concurrent Lambdas. }
	\label{ripple_cost}
\end{table}

\subsection{Automated Provisioning}
\label{sec:evaluation:accuracy}

\noindent{\bf{Performance estimation accuracy: }}We first examine Ripple's ability 
to correctly provision resources in a serverless cluster. We assume that jobs from 
all three applications are submitted, and for simplicity, each job 
is trying to maximize its performance, instead of specifying a particular deadline. 
As discussed in Section~\ref{sec:design}, Ripple will first launch two short-lived 
canary runs for each job, with different degrees of concurrency, apply SGD to infer 
their performance with all other concurrency degrees, and determine the configuration 
that minimizes execution time for each new job. 
Fig.~\ref{fig:error_perf}a shows the distribution of error between the performance estimated by Ripple 
using Stochastic Gradient Descent (SGD) and the actual performance measured on an 
AWS Lambda cluster when running with the corresponding degree of concurrency. 
Errors are low across all three applications, which allows Ripple to accurately 
determine the number of Lambdas per execution phase that will maximize an application's 
performance. The higher errors at the tails of the three violin plots correspond to 
the first few jobs that arrive, at which point Ripple's knowledge of 
application characteristics is limited, and the number of applications in the SGD matrix 
is small. As more applications arrive in the system, the estimation accuracy improves. 

Profiling and inference with SGD incur some scheduling overheads to each application. 
For the examined applications, profiling overheads range from 260ms for jobs with 1-3 stages 
to 6s for jobs with many stages, like the kNN classification. Inference overheads are 
less than 60ms in all cases. Aggregate overheads are always less than 4\% of the job's 
execution time. 


\noindent{\bf{Performance predictability: }}Fig.~\ref{fig:error_perf}b shows 
the distribution of execution time for 100 jobs of each application class 
with Ripple's provisioning mechanism, compared to (i) using the default 1MB 
data chunk size, and (ii) always using the maximum number of Lambdas that 
AWS allows (1,000 for our setting). Across all application classes Ripple 
achieves the best performance, especially in the case of kNN classification, 
where resource demands are high, and incorrect provisioning has a severe 
impact on execution time. More importantly, performance with Ripple is highly 
predictable, despite the framework not having control over where serverless 
functions are physically placed on AWS. This is in part because Ripple 
selects the number of Lambdas to be sufficient to exploit the job's parallelism, 
but not high enough to cause long queueing delays from AWS's resource 
limit, and in part because of Ripple's straggler detection technique. 
Straggler respawning reduces variability 
by respawning the few tasks that are underperforming due to interference or faulty data records. 

In comparison, execution time for the 1MB and maxLambdas policies 
varies widely and experiences long tails, especially when using 
the maximum concurrency allowed. There are two reasons for this. 
First, when using the default data chunk size 
with very large input datasets, as is the case with the examined applications, 
the number of total Lambdas that need to execute exceeds the total allowed limit 
by a lot, leading to long queueing delays, as functions wait for resources 
to become available. Even if AWS's resource limit increased, fine-grained 
parallelism does not come for free, incurring overheads for work partitioning, 
synchronization, and combining the per-function outputs. 
Second, when using the maximum number of allowed Lambdas with large datasets, 
e.g., 500GB, each function is still responsible for a large amount of data, 
333MB in this case. Given the strict time limit Lambdas have today, 
this causes several functions to time out and be respawned by AWS. 
Note that this is not a case of a function becoming a straggler, 
and Ripple's straggler respawning technique would only increase 
the system load further by respawning functions that will time out again. 

Table~\ref{ripple_cost} also shows the cost for each policy 
across applications. Ripple incurs the lowest cost, since it avoids 
excessive task queueing and respawning. The maxLambdas policy has the 
highest cost in the resource-intensive SpaceNet application which 
results in many Lambdas timing out and needing to rerun, while for 
the other two scenarios, the excessive number of Lambdas used by the 
1MB default split size policy incurs the highest cost, since the 
increased number of Lambdas does not also result in faster execution. 











\begin{figure}
\centering
\begin{tabular}{cc}
	\multicolumn{2}{c}{\includegraphics[scale=0.206,viewport=240 0 800 60]{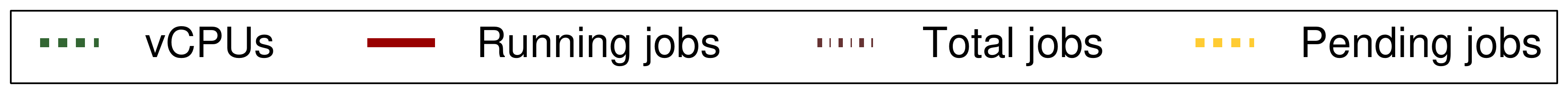}} \\
\includegraphics[scale=0.268,viewport=80 30 440 330]{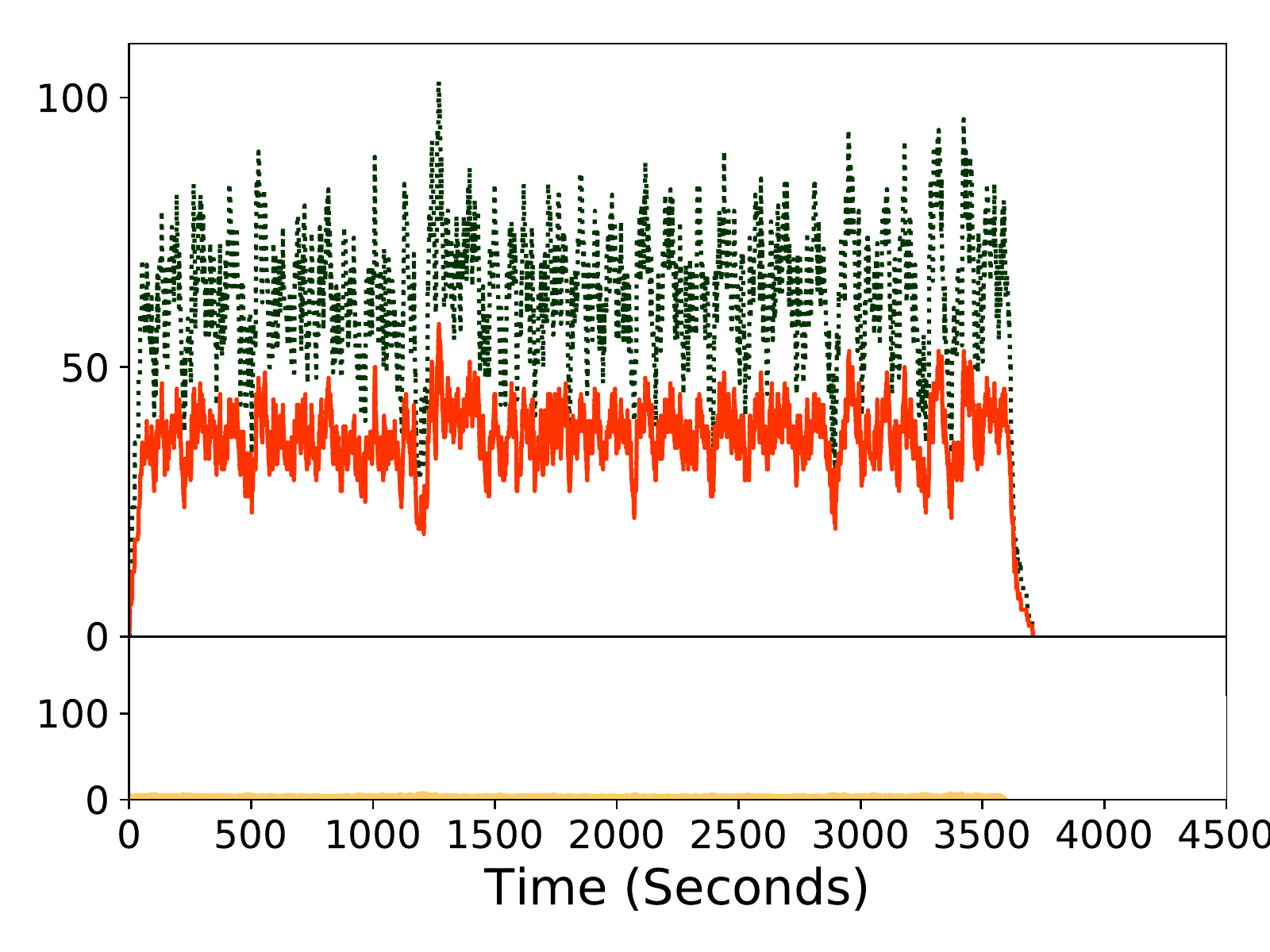} &
\includegraphics[scale=0.268,viewport=20 30 400 330]{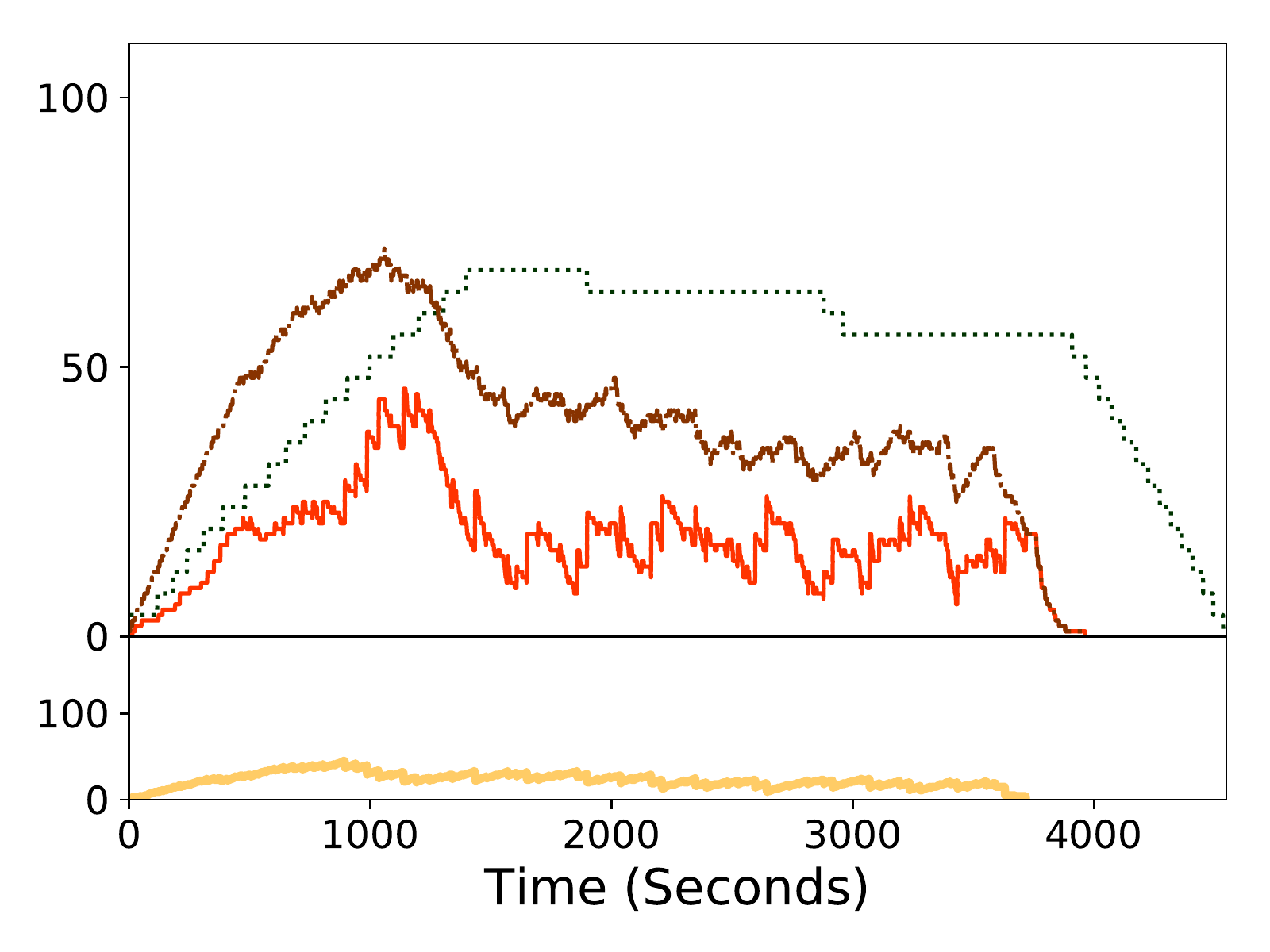} \\
{\footnotesize (a) Ripple} & {\footnotesize(b) Amazon EC2}
\end{tabular}
\caption{Tide performance under uniform load, with 1 job per 10s.
	(a) and (b) show the total vCPUs used (green), and the total (brown) and running jobs on the top (red).
In the bottom, we show the pending jobs waiting to run (yellow).}
\label{fig:tide_uniform_comparison}
\end{figure}

\subsection{Workload Distributions}
\label{sec:evaluation:workloads}

We now show Ripple's elasticity under different job arrival distributions, compared to Amazon EC2. 
We examine uniform, bursty, and diurnal load patterns. 

\begin{figure}
	\centering
	\begin{tabular}{cc}
	\multicolumn{2}{c}{\includegraphics[scale=0.206,viewport=240 0 800 60]{Legend.pdf}} \\
		\includegraphics[scale=0.268,viewport=40 30 440 330]{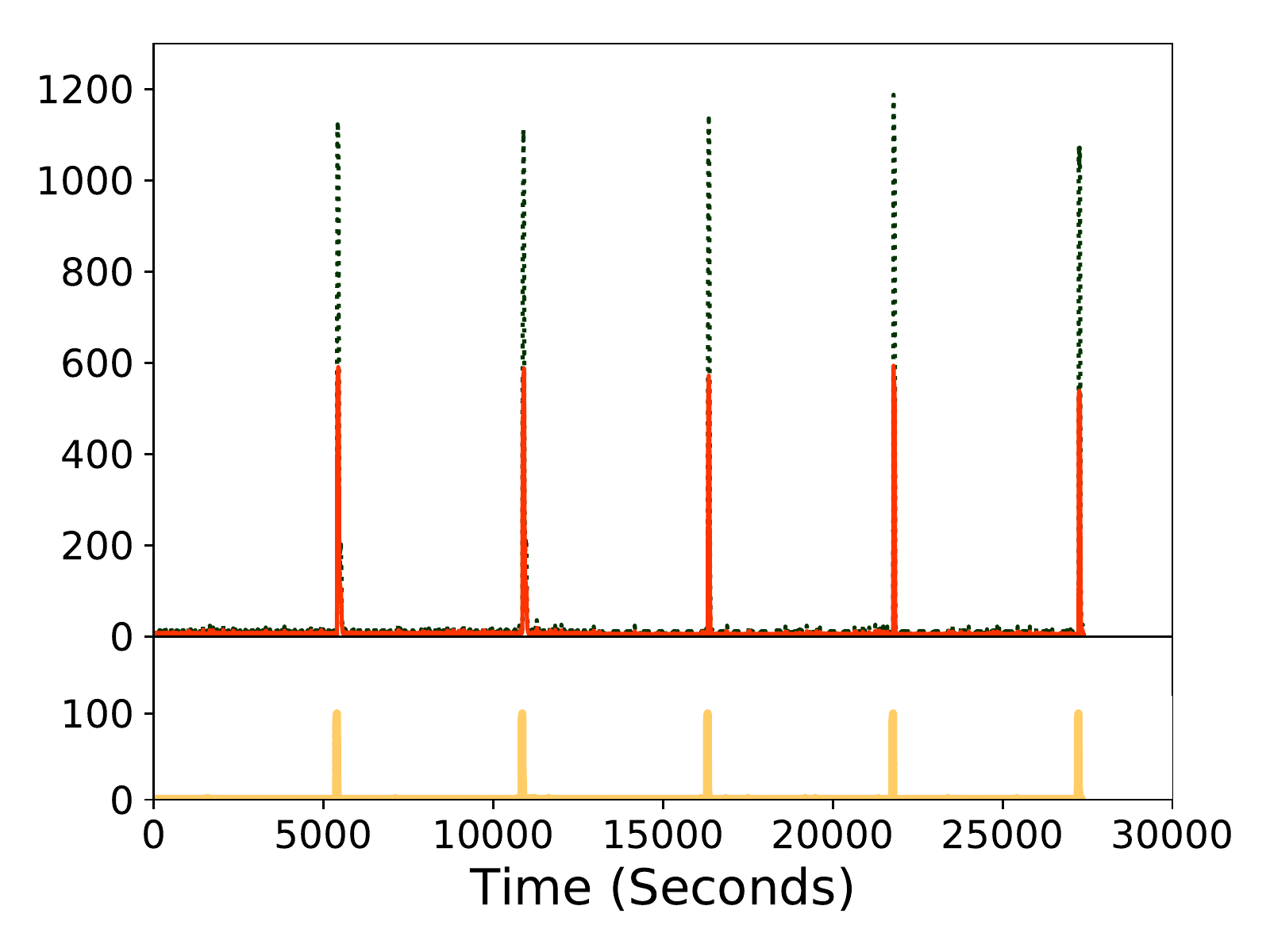} &
		\includegraphics[scale=0.268,viewport=30 30 440 330]{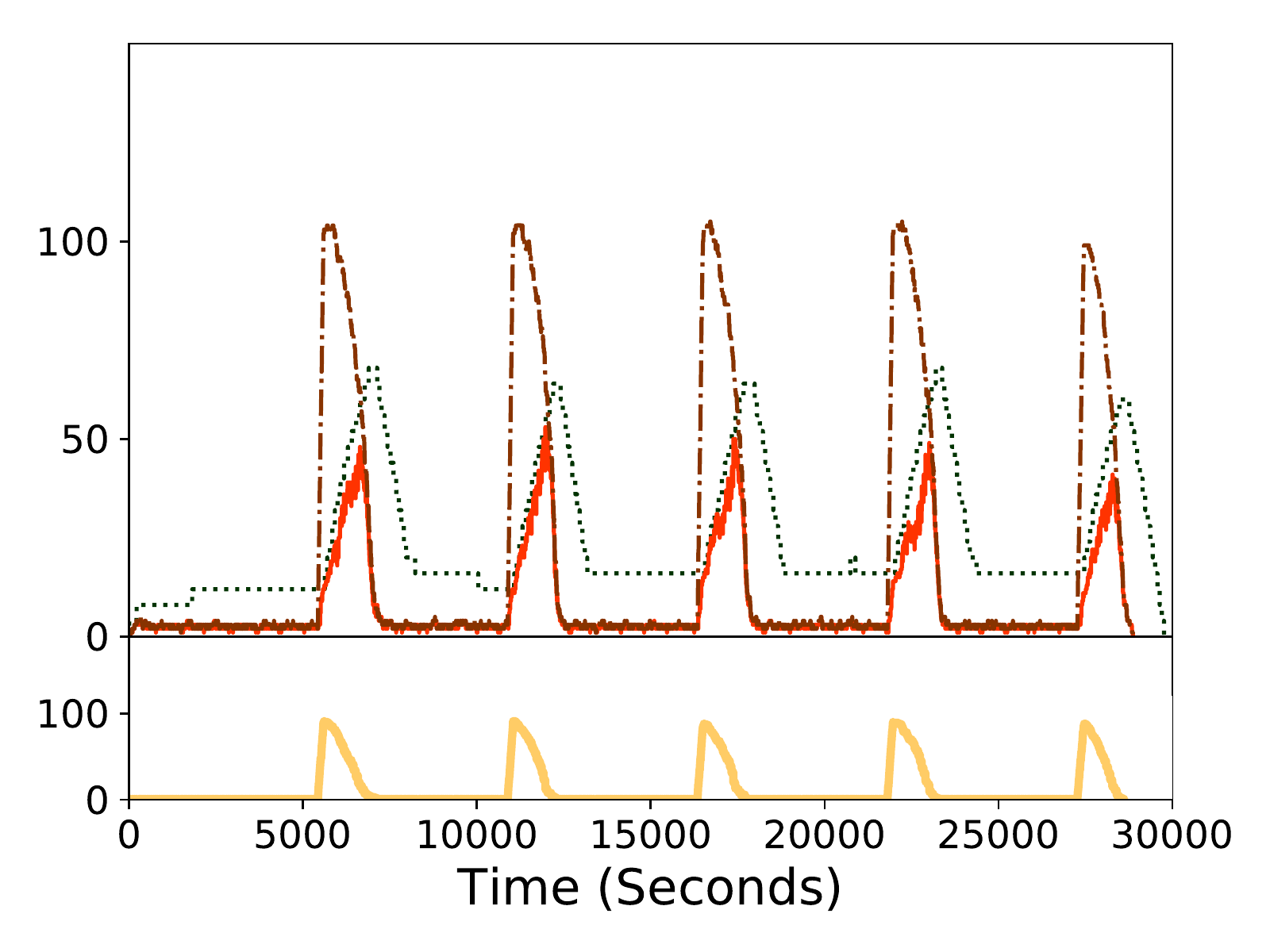} \\
		{\footnotesize (a) Ripple} & {\footnotesize (b) EC2}
	\end{tabular}
	\caption{Tide performance under bursty load (1 job/min). Every 90 min,
		a burst of 100 jobs arrives. (a) and (b) show the used vCPUs (green),
	and the total (brown) and running jobs (red). The bottom shows the pending jobs (yellow).}
	\label{fig:tide_bursty_comparison}
\end{figure}

\vspace{0.06in}
\noindent\textbf{Uniform: }
In the case of the Tide proteomics framework, we create a uniform workload by sending 
a new job every 10s for an hour. Figure~\ref{fig:tide_uniform_comparison} shows 
that both for EC2 and Lambda, due to the frequency of job arrivals, 
there is always at least a task running, with Ripple being able to immediately scale 
to meet the resource demands. EC2 was able to handle the input load once enough machines 
were launched, but as a result of the initial adjustment period, it took an additional 
6 minutes to finish the entire scenario. With Ripple, the average job completion time 
was $4.5\times$ faster than EC2, 2min on average, whereas the 
average job completion time for EC2 was approximately 8.5min. The cost for Lambda is also 
lower compared to EC2. 
Similarly, DNA compression achieves $8\times$ faster execution with Ripple compared to EC2 under a uniform load (85s with Ripple as opposed to 11min with EC2). 

\begin{figure}
   \centering
    \begin{tabular}{cc}
      \includegraphics[scale=0.268,viewport=60 30 440 330]{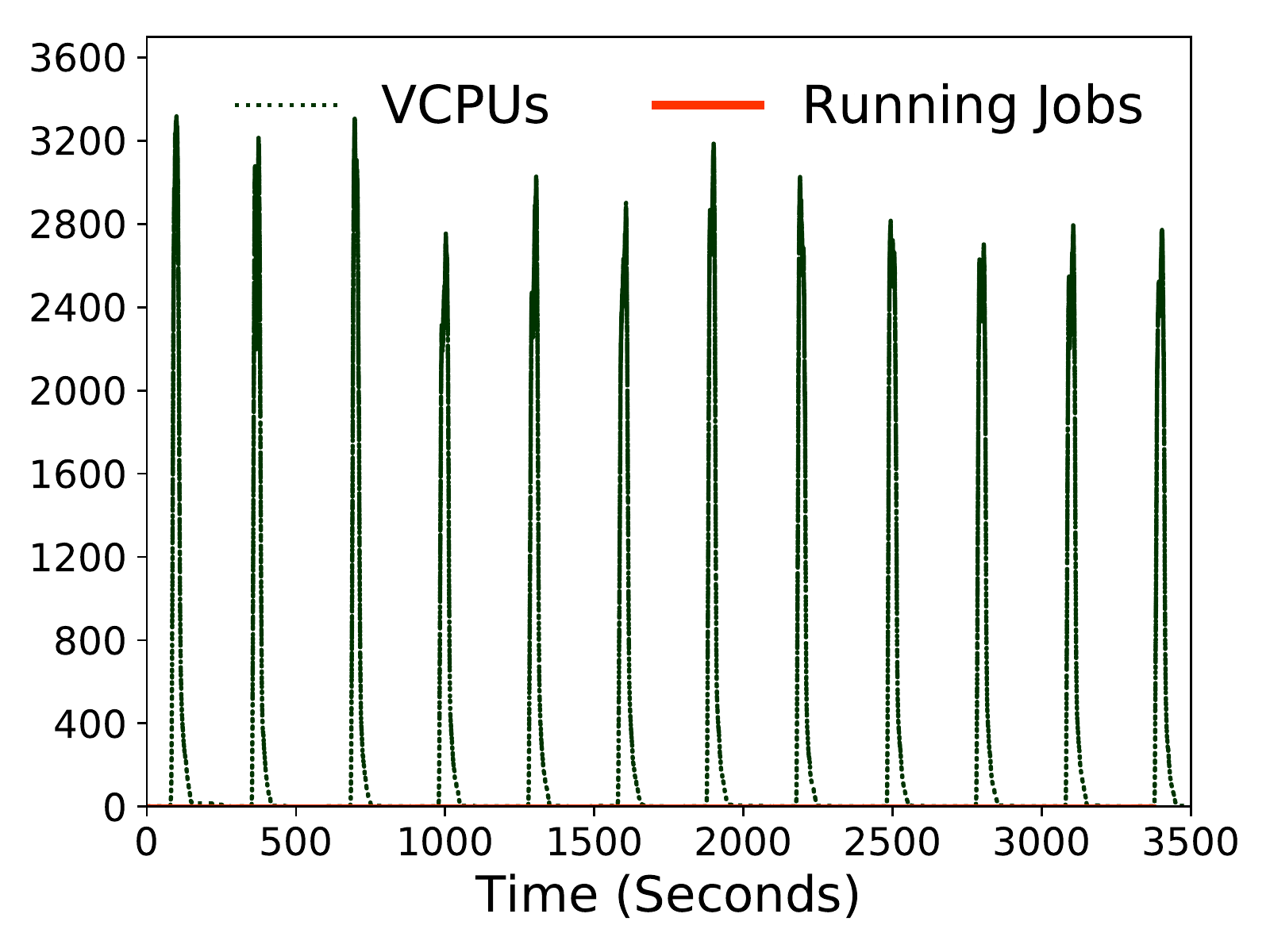} &
      \includegraphics[scale=0.268,viewport=30 30 440 330]{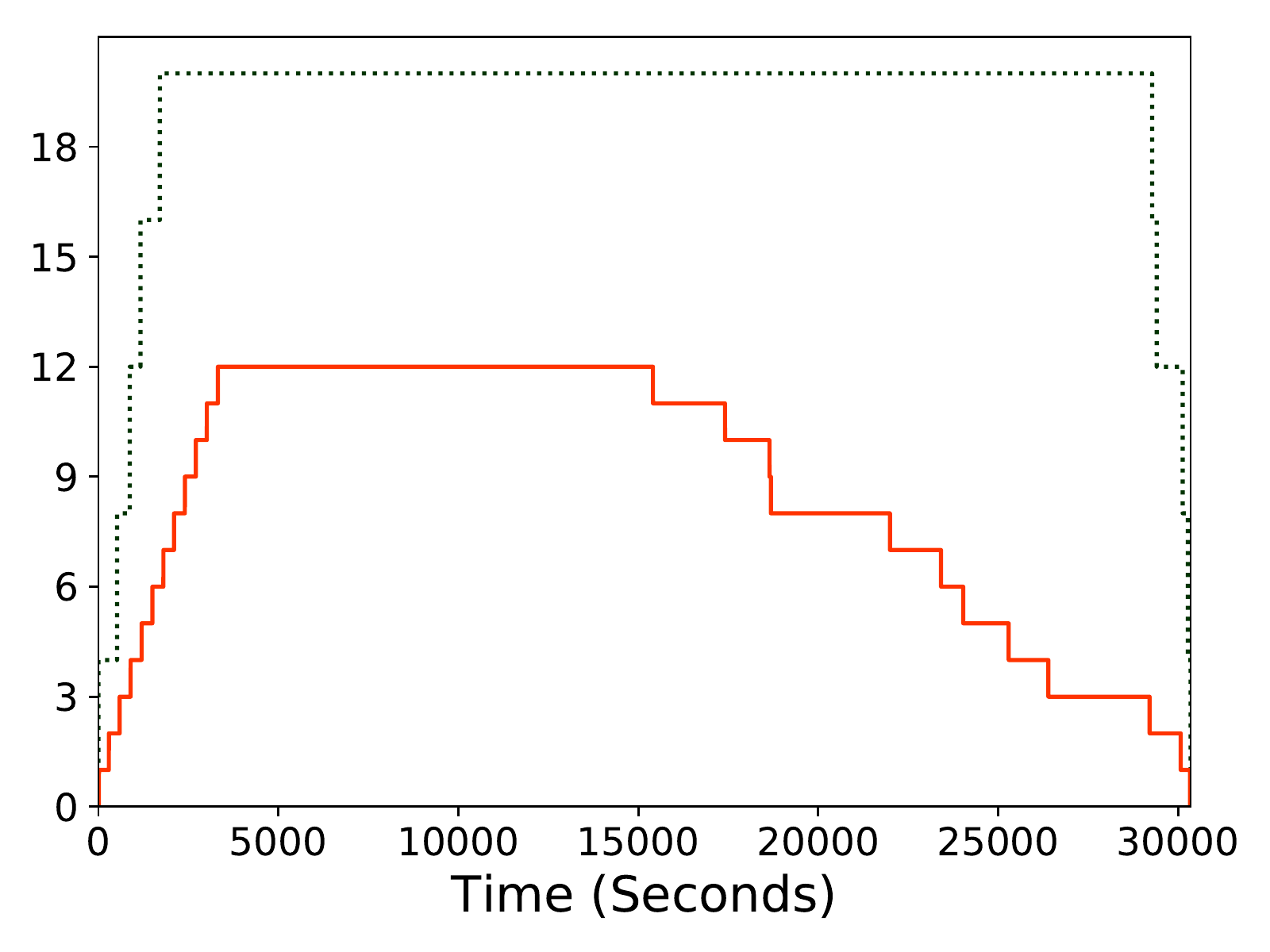} \\
      {\footnotesize (a) Ripple} & {\footnotesize (b) EC2}
    \end{tabular}
    \caption{SpaceNet performance under a uniform workload. Over an hour, 
    we send 1 job every 5min. Both graphs show the vCPUs used by running jobs (green), 
    as well as the total number of jobs actively processed (red).}
    \label{fig:spacenet_uniform_comparison}
\end{figure}

We also examine a uniform workload distribution for SpaceNet, where we sent one job 
every 5 minutes for an hour. Figure~\ref{fig:spacenet_uniform_comparison} shows the 
results for Ripple and EC2. Both Ripple and EC2 were immediately able to handle new 
incoming requests; however, due to the memory requirements of each classification job, 
the average job completion time for Ripple was more than an order of magnitude faster 
(i.e., $80\times$ faster) at 4.5 minutes, with EC2 requiring approximately 6 hours to 
complete an equal amount of work. 

\begin{figure}
	\centering
	\begin{tabular}{cc}
	\multicolumn{2}{c}{\includegraphics[scale=0.206,viewport=240 0 800 60]{Legend.pdf}} \\
		\includegraphics[scale=0.268,viewport=40 30 440 330]{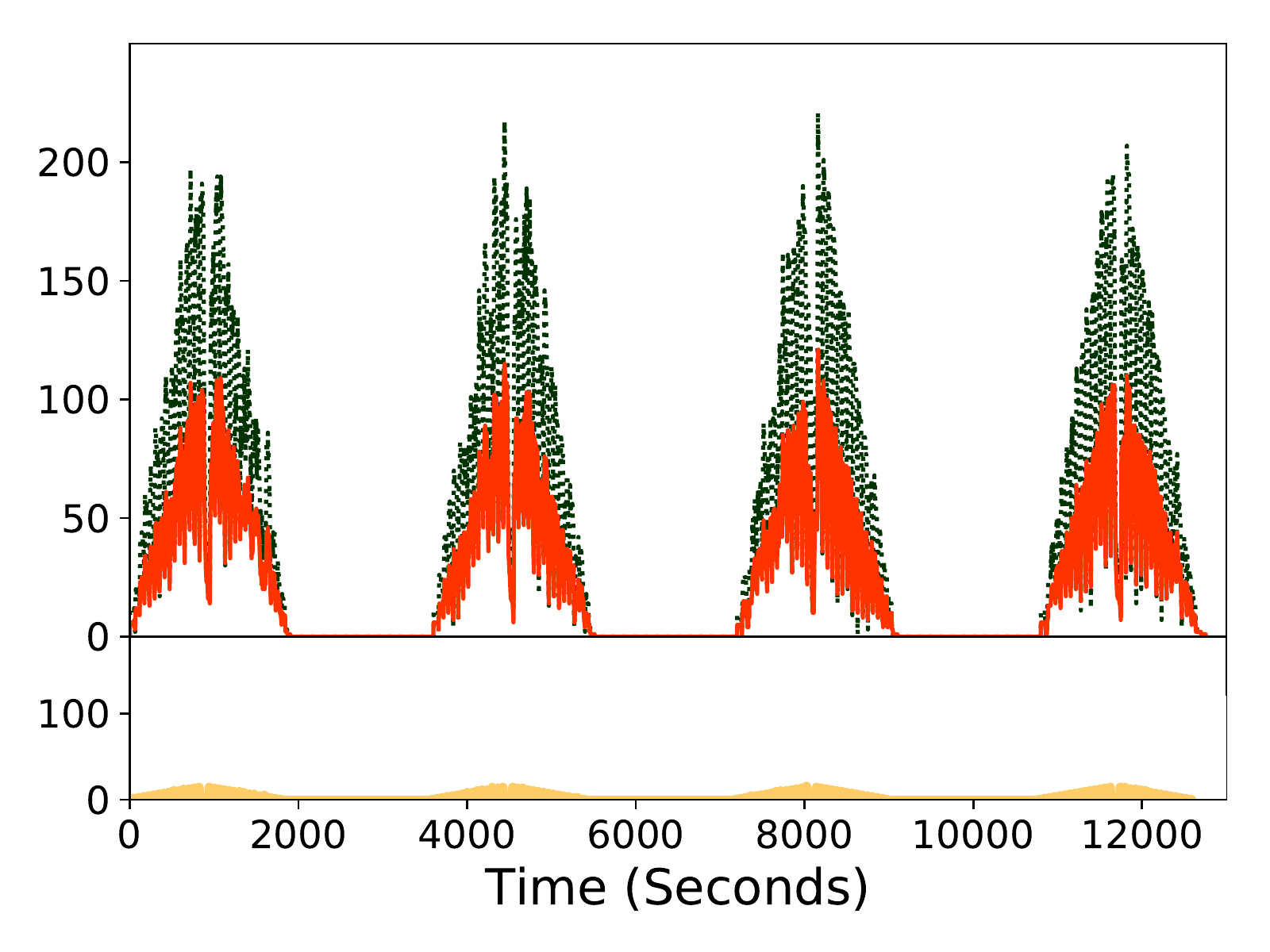} &
		\includegraphics[scale=0.268,viewport=30 30 440 330]{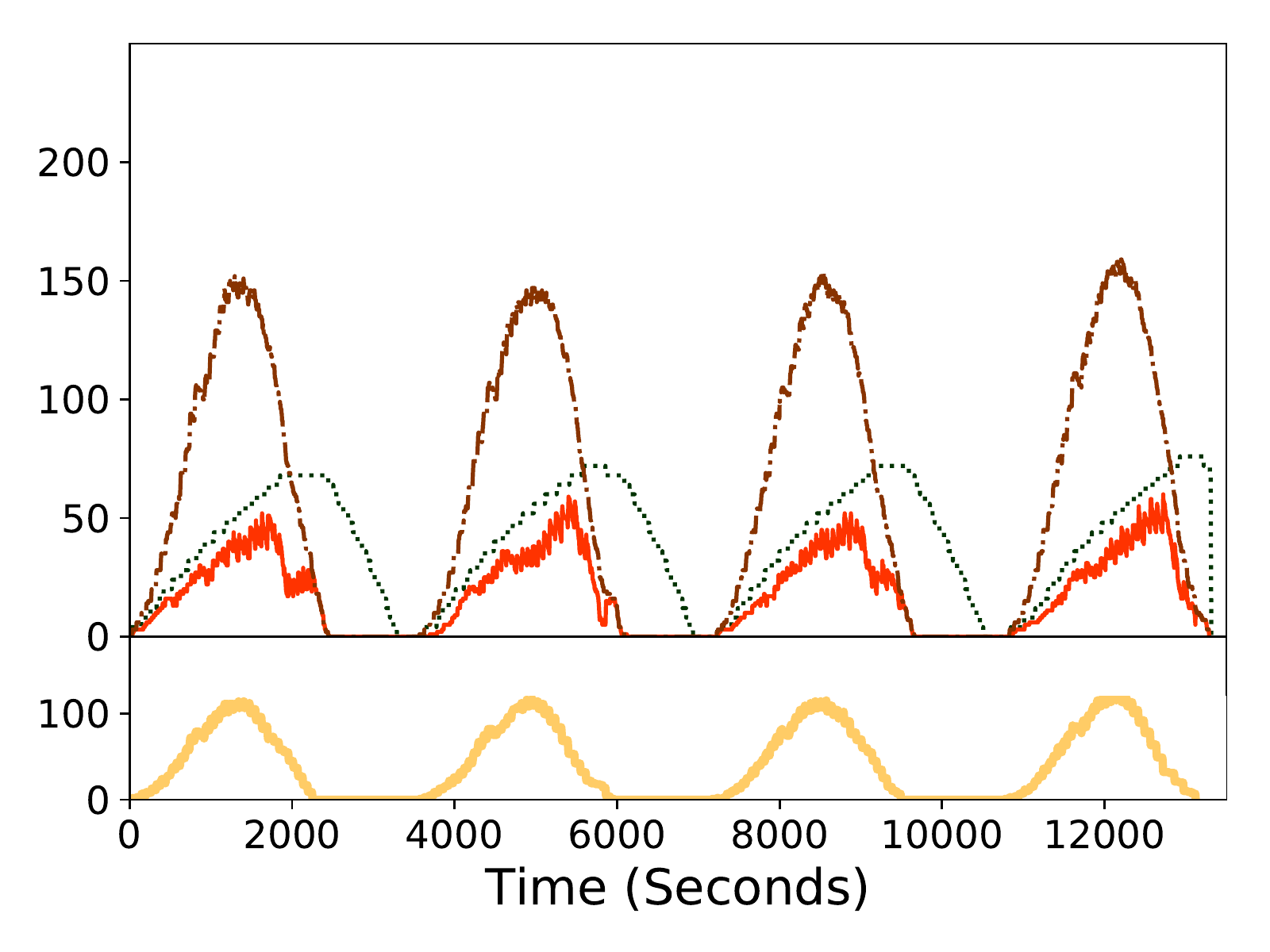} \\
		{\footnotesize (a) Ripple} & {\footnotesize (b) EC2}
	\end{tabular}
	\caption{Tide performance under diurnal load. Over 30min,
		        we increased the jobs from 0 to 15 and back to 0.
			(a) and (b) show the used vCPUs (green),
		and the total (brown) and running jobs (red). The bottom shows pending jobs (yellow).}
		\label{fig:tide_diurnal_comparison}
	\end{figure}

\vspace{0.06in}
\noindent\textbf{Bursty: }
We now initiate a Tide job every minute for 8 hours. Every 90min, we additionally send 
a burst of 100 jobs to emulate a sudden load spike. Figure~\ref{fig:tide_bursty_comparison} 
shows that, while the burst is happening, Ripple used 
almost 700 vCPUs for approximately 400 concurrent functions, which allowed it 
to immediately handle all 100 jobs with no performance degradation for any 
individual job. EC2, on the other hand, incurs the VM instantiation overhead, 
and hence takes a long time to scale resources to meet the requirements of all 
100 jobs. The figure shows that initially most jobs remained idle in the admission queue 
as they are waiting for more instances to be spawned. After the initial burst, 
EC2 did a better job at handling the burst; however, this was mostly because 
the machines were not given enough time to terminate during the non-bursty periods. 
Ripple, on the other hand, was able to handle the bursts by rapidly scaling up 
and down the number of active Lambdas, without incurring high costs 
by retaining unused resources during periods of low load. 
As a result, the average job completion time was $5\times$ faster for Ripple 
at about 2min, compared to 10min for EC2. 
The results are similar for DNA compression, with Ripple achieving $8\times$ faster execution compared to EC2. 

\vspace{0.06in}
\noindent\textbf{Diurnal:} Finally, we examine a diurnal job arrival distribution, 
which emulates the load fluctuation of many user-interactive cloud applications~\cite{Lo14,Lo15,Meisner11,Delimitrou12a}. 
During a 30 minute interval within a larger 4 hour period, we progressively 
increase the number of arriving jobs, and then progressively decrease back to zero new jobs. 
We repeat this pattern multiple times within the 4 hour period. 
Figure~\ref{fig:tide_diurnal_comparison} shows the diurnal results for Ripple and EC2. 
The results show that the average job completion time was $6.75\times$ faster 
for Ripple at about 2 minutes, whereas it was 13.5 minutes for EC2.


\begin{figure}
  \centering
  \begin{tabular}{cc}
    \includegraphics[scale=0.268,viewport=40 0 440 330]{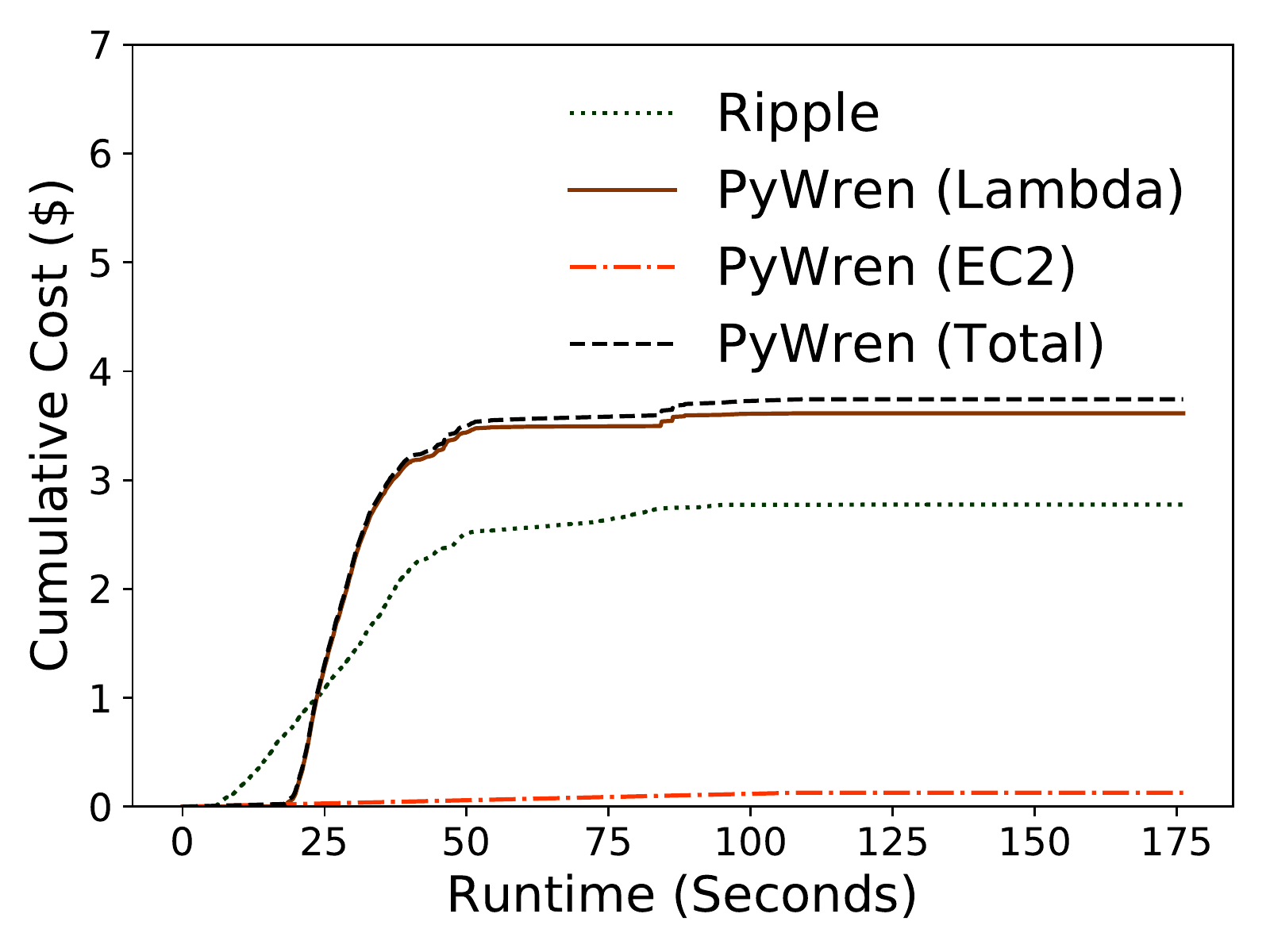} &
    \includegraphics[scale=0.268,viewport=30 0 440 330]{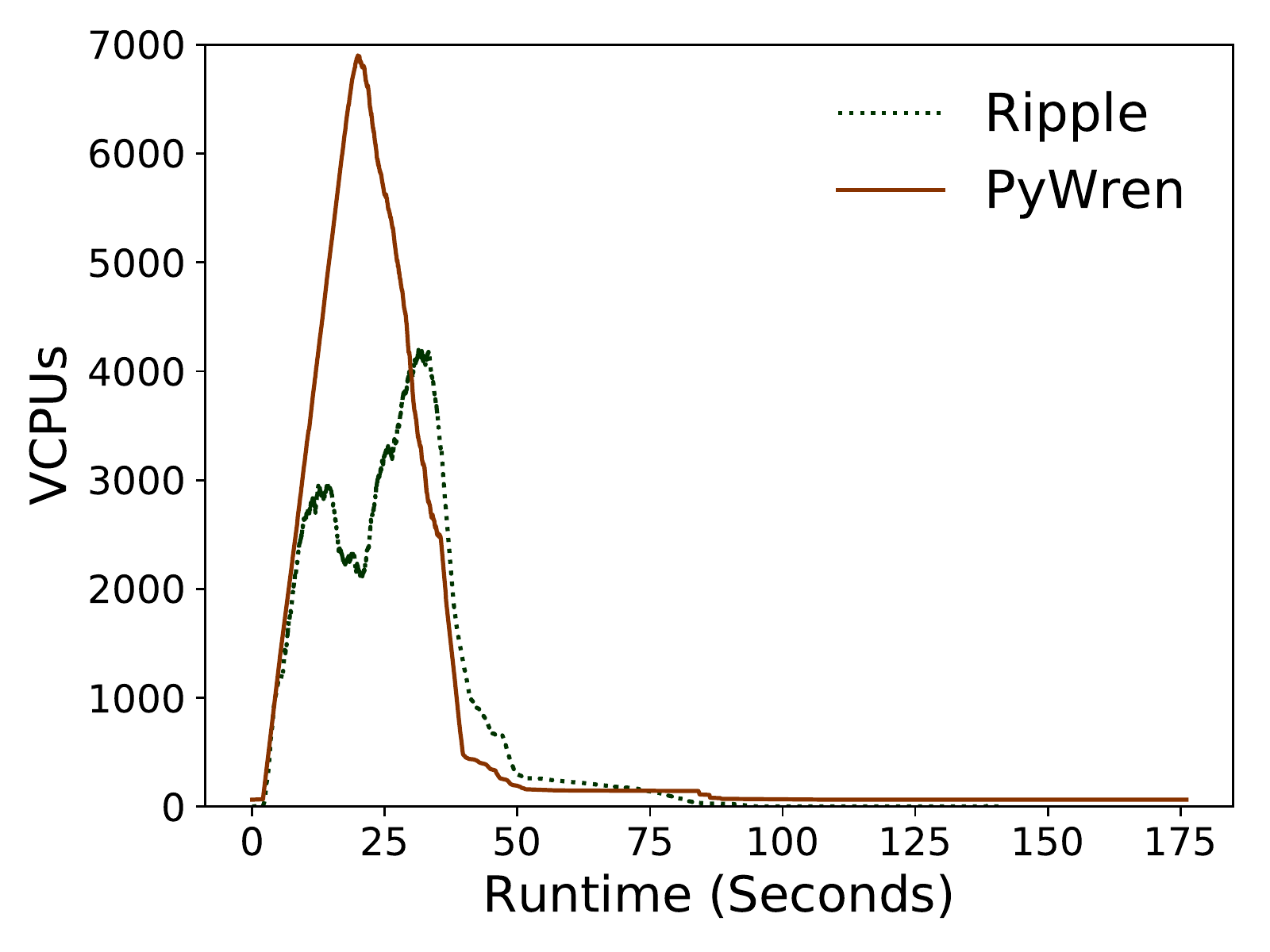} \\
  {\footnotesize (a) Cost Comparison} & {\footnotesize (b) vCPUs Comparison}
  \end{tabular}
  \caption{Comparison between PyWren and Ripple for SpaceNet: (a) shows the cumulative cost of Lambda for Ripple and PyWren, 
  as well as the cost of EC2 and overall cost for PyWren; 
  (b) shows a comparison of the number of vCPUs used at each point in time by the two frameworks.}
  \label{fig:pywren_comparison}
\end{figure}

\subsection{PyWren Comparison}
\label{sec:evaluation:pywren}

We now use the SpaceNet application to compare Ripple and PyWren. 
We send one SpaceNet job to each framework. To allow for the increased 
compute and memory requirements of SpaceNet, we perform this experiment 
in an AWS region where the max function concurrency was 5,000 Lambdas.
Figure~\ref{fig:pywren_comparison} shows a comparison of the number of 
vCPUs and the cumulative cost for each framework. The runtime for Ripple 
was 25.7\% faster at about 140 seconds, while the runtime for PyWren was 
about 176 seconds. Part of the reason is that Ripple can explicitly invoke 
the next stage of the pipeline, whereas stages in PyWren have to wait for 
S3 results. For SpaceNet, the maximum concurrency of a stage is 6,764 
functions. This meant PyWren had to launch almost 7K functions simultaneously. 
Because Ripple provisions each execution phase separately, it did not need 
7K functions for every stage of the pipeline, avoiding resource and cost inefficiencies. 
This translates to a \$2.77 cost for a single run for Ripple and \$3.61, for PyWren. 
The cost of the EC2 machine during this period was \$0.13. The rest of the 
cost difference was caused by the difference in the runtime of individual 
serverless functions. 

\begin{figure}
  \centering
  \begin{tabular}{cc}
	  \includegraphics[scale=0.268,viewport=40 0 440 290]{tide100.pdf} &
  \includegraphics[scale=0.268,viewport=30 0 440 290]{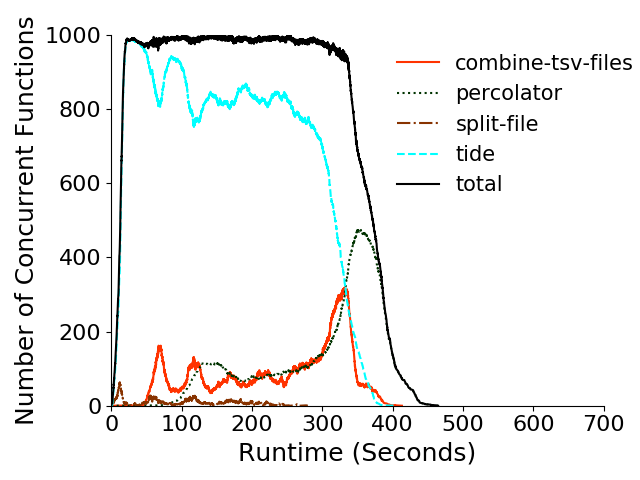} \\
  {\footnotesize (a) 100 concurrent} & {\footnotesize  (b) 1,000 concurrent}
  \end{tabular}
  \caption{Performance of 100 versus 1,000 concurrent Tide jobs. 
  The graphs show a breakdown of the number of Lambdas actively used 
  during each execution phase. }
  \label{fig:tide_concurrency}
\end{figure}

\subsection{Job Concurrency}
\label{sec:evaluation:concurrency}

We now evaluate how well Ripple scales with the number of concurrent jobs.
Figure~\ref{fig:tide_concurrency} shows the per-stage execution time breakdown 
for 100 concurrent Tide jobs versus 1,000 jobs.
100 concurrent jobs do not generate enough Lambdas to reach the function limit, 
while 1,000 concurrent jobs almost immediately hit this limit.
The fluctuation in the number of active Lambdas across phases for both concurrency 
levels is approximately the same, despite the higher concurrency experiment reaching 
the upper Lambda limit. 
The total runtime was about twice as long for 1,000 concurrent jobs and, in general, 
used twice the number of serverless functions. 




\subsection{Fault Tolerance}
\label{sec:evaluation:fault_tolerance}

\begin{figure}
\centering
\begin{tabular}{cc}
	\includegraphics[scale=0.268,viewport=50 0 400 320]{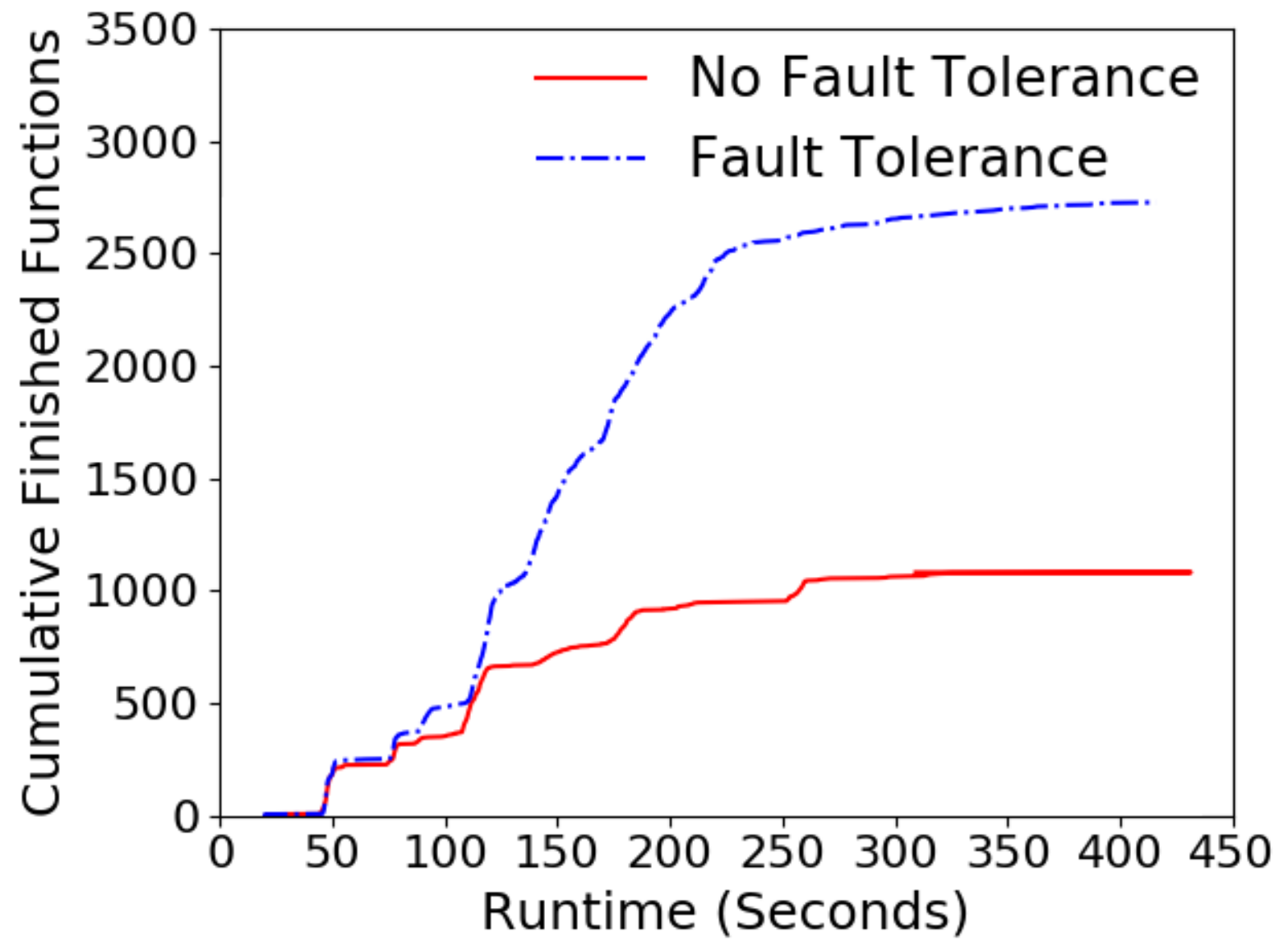} & 
  \includegraphics[scale=0.268,viewport=0 0 400 320]{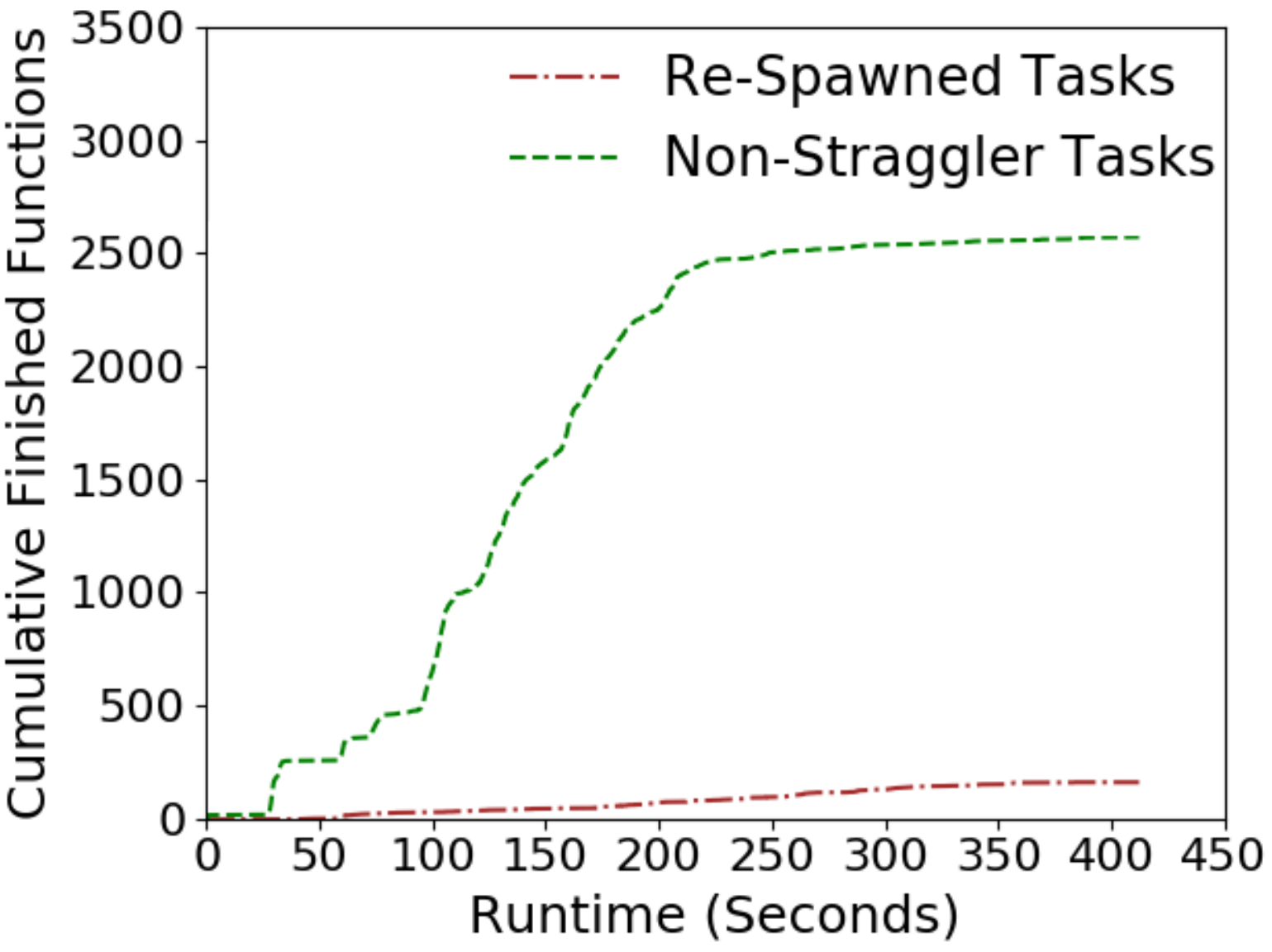}
  \end{tabular}
  \caption{(a) CDF of Lambda completion time for 20 DNA compression jobs, 
  with and without Ripple's fault tolerance. (b) Comparison between normal 
  and re-spawned tasks by Ripple. }
  \label{fig:fault_tolerance_comparison}
\end{figure}

We now quantify the effectiveness of Ripple's fault tolerance mechanism. 
We run 20 DNA compression jobs in parallel, and force a 10\% failure 
probability on each task. In the case of Ripple, under-performing tasks, 
i.e., tasks not making sufficient progress are proactively re-spawned to 
avoid performance degradation. Figure~\ref{fig:fault_tolerance_comparison} 
shows the impact of the fault tolerance mechanism on execution time, 
compared to the baseline AWS system. When no fault tolerance is used, 
only 4 jobs complete successfully without their Lambdas reaching the 
5 minute timeout limit, and being terminated. Instead, when Ripple 
proactively detects and re-spawns stragglers, all jobs complete 
within AWS Lambda's time limit, as seen in the CDF of Fig.~\ref{fig:fault_tolerance_comparison}a. 
Fig.~\ref{fig:fault_tolerance_comparison}b shows the tasks that had 
to be re-spawned by Ripple to avoid degrading the overall job's execution time. 

%% file: related_work.tex
\section{Related Work}
\label{sec:related}

\subsection{Programming Frameworks for Serverless}

The past few years a number of programming frameworks for serverless 
have been designed that either target specific applications, or focus on generality. 

\vspace{0.03in}
\textbf{PyWren: }PyWren targets general-purpose computation, 
and offers a simple programming framework in Python that evaluates 
the viability of serverless for distributed computing applications~\cite{pywren}.
The authors show that embarrassingly parallel workflows, such as computational imaging, 
solar physics, and object recognition, can be implemented in a straightforward way using AWS Lambda.
PyWren divides computation into \textit{map} and \textit{reduce} phases, similar 
to the MapReduce model~\cite{Mapreduce}, with mappers using Lambdas to run in parallel, 
while reducers rely on long-running EC2 instances for serial computation. Locus~\cite{locus} extends PyWren, 
and uses a mixture of S3 and Redis to scale sorting on Lambda.

PyWren highlights the potential of serverless for jobs with a single parallel phase and 
computationally-expensive serial phases, such as \textit{reduce}, which can run on a long-running EC2 instance. 
Unfortunately the fact that PyWren still relies on long-running instances
for serial execution phases reduces the performance and cost benefits of serverless.
Additionally, since PyWren uses both serverless and EC2 for the job's workflow, 
cloud maintenance remains complicated for users who must manage both serverless and on-demand resources. 
Finally, since PyWren provisions Lambdas for the entire job once, it cannot adjust to the different degrees of 
parallelism different computation phases often have, hurting resource efficiency, and limiting the complexity of supported tasks. 

\vspace{0.03in}
\textbf{EMR Cluster: }EMR Cluster dynamically adjusts the size of a cluster depending on a user's resource needs~\cite{emr}.
Spark can run on top of EMR Cluster, and users are charged on a per-second granularity.
However, it can take several minutes to launch a new machine in an EMR Cluster, 
unlike serverless, where functions are initialized in under a second.

\vspace{0.03in}
\textbf{mu: }\texttt{mu} is a serverless framework targeting a specific class of applications, namely video processing, 
and is used by the ExCamera video encoding service~\cite{xcamera}. \texttt{mu} uses {\textsc AWS} Lambda
for both map and reduce tasks, each framed as a short-lived worker.
In addition to AWS Lambda, \texttt{mu} requires a long-lived {\textsc EC2} 
server for the Coordinator, and another long-lived {\textsc EC2} server to act as the Rendezvous Server.
The Coordinator assigns tasks to workers and tracks application state, 
and the Rendezvous Server passes messages between workers. 

\vspace{0.03in}
\textbf{GG: }\texttt{gg} is a serverless framework where dependencies between tasks, such as distributed compilation, 
are explicitly known in advance. Similar to \texttt{mu}, \texttt{gg} exploits serverless to parallelize interdependent 
tasks~\cite{thunk}, showing significant speedups for computation with irregular dependencies, such as software compilation.
However, \texttt{gg} requires the dependency tree for a task to be pre-computed, making it impractical for cases where 
job dependencies are not known in advance, or when dependencies depend on the input. 

\vspace{0.03in}
\textbf{SAND: }{\textsc{SAND}} is a framework designed to simplify interactions between serverless functions~\cite{sand}.
Currently, public clouds do not provide a way for the user to express 
how functions interact with each other, optimizing instead 
for individual function execution. 
{\textsc SAND} co-locates functions from the same pipeline 
on the same physical machine, and reuses the same containers 
for a pipeline's functions. While beneficial for performance 
and resource utilization, {\textsc SAND} assumes full 
cluster control to enforce function placement to machines, which is not possible in public cloud providers. 

\vspace{0.03in}
\textbf{Pocket: }The short execution times of Lambda functions require fast communication between functions and storage. 
Like Locus, Pocket addresses this by dynamically scaling storage 
to meet a given performance and cost constraint~\cite{pocket, Klimovic18}. 
Pocket relies on user-provided hints on the performance criticality 
of different jobs, their storage requirements, and their provisioning 
needs in terms of Lambda functions. 

While all these frameworks highlight the increasing importance of developing systems for serverless compute, 
they currently do not provide the automated resource provisioning, scheduling, and fault tolerance mechanisms 
cloud applications need to meet their QoS requirements. 

\vspace{-0.05in}
\subsection{Dataflow Frameworks}

Dataflow frameworks have been an active area of research in both academic and industrial settings for over 15 years, particularly in the context of analytics and ML applications.

\vspace{0.03in}
\textbf{Analytics Frameworks: }MapReduce~\cite{Mapreduce} innovated using parallel, stateless dataflow computation 
for large analytics jobs, while Spark~\cite{spark} focused on improving the performance and fault tolerance 
of interactive and iterative jobs by adding in-memory caching to persist state across computation phases. 
DryadLINQ~\cite{dryadlinq} is another example of a framework that provided abstractions on top of a MapReduce-like system, 
enabling developers to write high-level programs, which the system transparently compiles into distributed computation. 

\vspace{0.03in}
\textbf{ML Frameworks: }GraphLab~\cite{graphlab} and Naiad~\cite{naiad} are both programming frameworks optimized for ML applications. 
DistBelief~\cite{distbelief}, TensorFlow~\cite{tensorflow}, and MXNet~\cite{mxnet} provide similar capabilities, 
focusing on deep neural networks. All three provide higher-level APIs for defining dataflows 
which consist of stateless, independent workers and stateful servers for sharing global parameters. 
TensorFlow's dataflow graph is static; unlike Ripple, it cannot handle dynamically-generated computation 
graphs. Ray~\cite{ray} enables dynamic graphs, but its design is specific to reinforcement learning pipelines, 
while Ripple targets general parallel computation. 
\vspace{-0.05in}




%% file: conclusion.tex
\section{Conclusion}
\label{sec:conclusions}

We presented Ripple, a high-level declarative programming framework for serverless compute. 
Ripple allows users to express the dataflow of complex pipelines at a high-level, and handles work and data 
partitioning, task scheduling, fault tolerance, and provisioning automatically, 
abstracting a lot of the low-level system complexity from the user. We port three large ML, 
genomics, and proteomics applications to Ripple, and show significant performance benefits compared 
to traditional cloud deployments. We also show that Ripple is able to accurately determine the degree of concurrency needed 
for an application to meet its performance requirements, and further improves performance predictability by 
detecting straggler tasks eagerly and respawning them. Finally, we showed that Ripple is robust across 
different job arrival distributions and different degrees of concurrency, and implements 
a number of scheduling policies, including round-robin, priorities, and deadline-based scheduling. 

\section*{Code Availability}

Ripple is open-source software under a GPL license (\url{https://github.com/delimitrou/Ripple}), and is already in use by several research groups. 